\documentclass[lettersize,journal,doublecolumn]{IEEEtran}%
\usepackage{amsmath,amsfonts}
\usepackage{algorithm}
\usepackage{array}
\usepackage{textcomp}
\usepackage{stfloats}
\usepackage{url}
\usepackage{verbatim}
\usepackage{graphicx}
\usepackage{cite}
\usepackage{balance}
\usepackage[
font=small,labelfont=bf
]{caption}
\usepackage{amsfonts, amssymb, amsthm, color,mathtools}
\usepackage{bbm}
\usepackage{comment,enumitem}

\usepackage{tikz}
\usepackage{diagbox}
\usepackage[capitalize]{cleveref}
\usepackage{caption}
\usepackage{subcaption}
\usepackage{dsfont}
\usepackage[nolist]{acronym}
\usepackage{mathtools} %
\usepackage[scientific-notation=true]{siunitx}
\usepackage{multicol}
\usepackage{multirow}

\usepackage[normalem]{ulem}
\usepackage{algpseudocode}
\algnewcommand{\Initialize}{%
  \State \textbf{Initialize:}
}

\newcommand{\imgscale}{0.7}

\newcommand{\sq}[1]{\ensuremath{\left[#1\right]}}
\newcommand{\curly}[1]{\ensuremath{\left\{#1\right\}}}
\newcommand{\br}[1]{\ensuremath{\left(#1\right)}}
\newcommand{\rv}[1]{\ensuremath{\texttt{#1}}}
\newcommand{\kl}[2]{\ensuremath{\text{D}_{\text{KL}}\left(#1 \Vert #2\right)}}
\newcommand{\klabbrev}[2]{\ensuremath{D_{\text{KL}}\left(#1 \Vert #2\right)}}
\newcommand{\klminus}[2]{\ensuremath{\text{D}_{\text{KL}}^-\left(#1 \Vert #2\right)}}
\newcommand{\klmin}{\ensuremath{\text{D}_{\text{KL}, \min}}}
\newcommand{\klmax}{\ensuremath{\text{D}_{\text{KL}, \max}}}
\newcommand{\klepsmin}{\ensuremath{\text{D}_{\text{KL}\epsilon, \min}}}

\newcommand{\card}[1]{\ensuremath{\left\vert#1\right\vert}}

\newcommand{\ceil}[1]{\ensuremath{\left\lceil #1 \right\rceil}}

\newcommand{\argmin}{\ensuremath{\text{arg}\min}}

\newcommand{\ps}[1]{\ensuremath{\mathcal{P}\br{#1}}}
\newcommand{\define}{\ensuremath{\coloneqq}}

\newcommand{\mgfarg}{\ensuremath{\phi}}
\newcommand{\samplesize}{\ensuremath{d}}
\newcommand{\samplecnt}{\ensuremath{m}}
\newcommand{\sampleidx}{\ensuremath{\ell}}

\newcommand{\samplevec}{\ensuremath{\mathbf{x}_\sampleidx}}
\newcommand{\samplematrix}{\ensuremath{\mathbf{X}}}
\newcommand{\workermatrix}[1][\armidx]{\ensuremath{\samplematrix_{#1, \timeidx}}}
\newcommand{\samplelabel}{\ensuremath{y_\sampleidx}}
\newcommand{\samplelabelvec}{\ensuremath{\mathbf{y}}}

\newcommand{\workervec}{\ensuremath{\samplelabelvec_{\armidx, \timeidx}}}
\newcommand{\model}{\ensuremath{\mathbf{w}}}
\newcommand{\modelopt}{\ensuremath{\model_\star}}

\newcommand{\initvec}{\ensuremath{\model_0}}
\newcommand{\lossfct}[1][]{\ensuremath{F(\samplematrix, \samplelabelvec, \model_{#1})}}
\newcommand{\workerlossfct}[1][]{\ensuremath{F(\workermatrix, \workervec, \model_{#1})}}
\newcommand{\lossfctsample}[1][]{\ensuremath{F(\samplevec, \samplelabel, \model_{#1})}}
\newcommand{\lr}{\ensuremath{\eta}}
\newcommand{\lipschitz}{\ensuremath{L}}
\newcommand{\gradvar}{\ensuremath{\sigma^2}}
\newcommand{\convexity}{\ensuremath{c}}
\newcommand{\batchsize}{\ensuremath{s}}
\newcommand{\batchcnt}{\ensuremath{k}}

\newcommand{\totalbudget}{\ensuremath{B}}
\newcommand{\workercnt}{\ensuremath{n}}
\newcommand{\budget}{\ensuremath{b}}
\newcommand{\armcnt}{\ensuremath{\workercnt}}
\newcommand{\armidx}{\ensuremath{i}}
\newcommand{\timeidx}{\ensuremath{j}}
\newcommand{\timevar}{\ensuremath{t}}
\newcommand{\iter}{\ensuremath{j}}

\newcommand{\timehorizon}{\ensuremath{T}}
\newcommand{\roundidx}{\ensuremath{r}}
\newcommand{\timehorizonround}[1][\roundidx]{\ensuremath{\timehorizon_{#1}}}
\newcommand{\respvar}{\ensuremath{Z}}

\newcommand{\armresp}[1][\timeidx]{\ensuremath{\respvar_\armidx^{#1}}}
\newcommand{\rvtmp}{\ensuremath{\rv{Z}}}
\newcommand{\element}{\ensuremath{\nu}}
\newcommand{\superarmvar}{\ensuremath{\mathcal{A}}}
\newcommand{\superarm}[1][\timeidx]{\ensuremath{\superarmvar^{\roundidx}(#1)}}
\newcommand{\anysuperarm}[1][\roundidx]{\ensuremath{\superarmvar^{#1}}}

\newcommand{\superarmelement}[1][\timeidx]{\ensuremath{\superarmvar^\roundidx}_\element (#1)}
\newcommand{\superarmbest}{\ensuremath{\superarmvar^{\roundidx, \star}}}
\newcommand{\superarmbestelement}{\ensuremath{\superarmvar^{\roundidx, \star}_\element}}
\newcommand{\superarmspace}[1][\roundidx]{\ensuremath{\mathcal{W}^{#1}}}
\newcommand{\superarmspacecardlim}{\ensuremath{\mathcal{W}^{\leq\cardmax}}}
\newcommand{\superarmresp}{\ensuremath{\respvar_{\superarm}^{\timeidx}}}
\newcommand{\superarmbestresp}[1][\timeidx]{\ensuremath{\respvar_{\superarmbest}^{#1}}}
\newcommand{\tmpvar}{\ensuremath{y}}
\newcommand{\tmpvarsecond}{\ensuremath{u}}
\newcommand{\tmpvarthird}{\ensuremath{p}}
\newcommand{\tmpvarforth}{\ensuremath{\xi}}
\newcommand{\tmpset}{\ensuremath{\mathcal{U}}}
\newcommand{\tmpsetsecond}{\ensuremath{\mathcal{I}}}
\newcommand{\tmpsetthird}{\ensuremath{\mathcal{S}}}

\newcommand{\iterworkers}{\ensuremath{\mathcal{A}(\iter)}}
\newcommand{\nonstragglers}{\ensuremath{\mathcal{R}(\iter)}}

\newcommand{\E}[1]{\ensuremath{\mathbb{E}\sq{#1}}}
\newcommand{\Var}[1]{\ensuremath{\textup{Var}\sq{#1}}}
\newcommand{\Prob}[1]{\ensuremath{\Pr\br{#1}}}

\newcommand{\cb}{\ensuremath{\text{LCB}}}

\newcommand{\cblong}{\ensuremath{\muihat(\timeidx-1) - \cbradshort{\armidx}{\timeidx-1}}}

\newcommand{\mui}{\ensuremath{\mu_\armidx}}
\newcommand{\muihat}{\ensuremath{\hat{\mu}_\armidx}}

\newcommand{\muhat}[1]{\ensuremath{\hat{\mu}_{#1}(\timeidx)}}
\newcommand{\mus}{\ensuremath{\mu_{\superarm}}}

\newcommand{\musbest}{\ensuremath{\mu_{\superarmbest}}}

\newcommand{\supergaplong}{\ensuremath{\mus - \musbest}}
\newcommand{\supergap}{\ensuremath{\Delta_{\superarm}}}

\newcommand{\supergapmax}[1][\roundidx]{\ensuremath{\Delta_{\anysuperarm[#1], \max}}}
\newcommand{\cbscale}{\ensuremath{f(\timeidx)}}
\newcommand{\cbradcomb}[1]{\ensuremath{\sqrt{\frac{2 \br{\roundidx+1} \log(\timeidx)}{#1}}}}
\newcommand{\cbradshort}[2]{\ensuremath{\theta_{#1}\br{#2}}}
\newcommand{\cbshort}[2][\timeidx]{\ensuremath{\cb_{#2}(#1)}}

\newcommand{\cardmax}{\ensuremath{e}}
\newcommand{\failcountermax}[1][\timeidx]{\ensuremath{C_{\armidx, \cardmax}\br{#1}}}

\newcommand{\pullcounter}[1][\timeidx]{\ensuremath{T_\armidx\br{#1}}}
\newcommand{\respsum}[1][\timeidx]{\ensuremath{M_\armidx\br{#1}}}
\newcommand{\pullcountervar}[2][\timeidx]{\ensuremath{T_{#2}\!\br{#1}}}

\newcommand{\counteroffset}{\ensuremath{h}}

\newcommand{\identity}[1]{\ensuremath{\mathds{1}\{{#1}\}}}

\newcommand{\expdev}[2]{\ensuremath{E\br{#1, #2}}}
\newcommand{\policy}{\ensuremath{\pi}}
\newcommand{\policyopt}{\ensuremath{\pi^\star}}
\newcommand{\cmabpolicy}{\ensuremath{\policy_{\text{cr}}}}
\newcommand{\klpolicy}{\ensuremath{\policy_{\text{kl}}}}

\newcommand{\regret}{\ensuremath{R_\iter^\policy}}

\newcommand{\subgamma}[2]{\ensuremath{\text{Sub}\Gamma\br{#1, #2}}}
\newcommand{\subgaussian}[1]{\ensuremath{\text{SubG}\br{#1}}}

\newcommand{\master}{main node}
\newcommand{\tmpa}{\ensuremath{\kappa}}
\newcommand{\tmpb}{\ensuremath{\tau}}

\newif\ifdouble
\doubletrue %

\ifdouble
\renewcommand{\imgscale}{1} \fi

\newtheorem{theorem}{Theorem}
\newtheorem{proposition}{Proposition}

\newtheorem{lemma}[theorem]{Lemma}

\newtheorem{example}{Example}
\newtheorem{remark}{Remark}

\theoremstyle{definition}
\definecolor{britishracinggreen}{rgb}{0.0, 0.26, 0.15}
\newtheorem{definition}{Definition}

\begin{document}

\title{Cost-Efficient Distributed Learning via Combinatorial Multi-Armed Bandits}

\author{%
  \IEEEauthorblockN{\textbf{Maximilian Egger}\IEEEauthorrefmark{1},
                    \textbf{Rawad Bitar}\IEEEauthorrefmark{1},
                    \textbf{Antonia Wachter-Zeh}\IEEEauthorrefmark{1},
                    and \textbf{Deniz Gündüz}\IEEEauthorrefmark{2}}\\
                    
  \IEEEauthorblockA{\IEEEauthorrefmark{1}%
                     Technical University of Munich,
                    Germany,
                    \{maximilian.egger, antonia.wachter-zeh, rawad.bitar\}@tum.de}\\
                    
  \IEEEauthorblockA{\IEEEauthorrefmark{2}%
                    Imperial College London,
                    United Kingdom,
                    \{d.gunduz\}@imperial.ac.uk}\\
\thanks{This project has received funding from a TUM-ICL JADS project (COALESCENCE), the German Research Foundation (DFG) under Grant no. WA3907/7-1 and was partly supported by the TUM - Institute for Advanced Studies, funded by the German Excellence Initiative
and European Union Seventh Framework Programme under Grant Agreement
No. 291763. D. Gündüz received support from European Research Council (ERC) through project BEACON (No. 677854).}%
\thanks{This work is currently under review for possible publication in IEEE Journal on Selected Areas in Communications.}%
\thanks{}}

\markboth{}%
{Egger \MakeLowercase{\textit{et al.}}: Efficient Distributed Machine Learning via Combinatorial Multi-Armed Bandits}

\begin{acronym}
\acro{sgd}[SGD]{stochastic gradient descent}
\acro{mab}[MAB]{multi-armed bandit}
\acrodefplural{mab}[MABs]{multi-armed bandits}
\acro{ucb}[UCB]{upper confidence bound}
\acrodefplural{ucb}[UCBs]{upper confidence bounds}
\acro{lcb}[LCB]{lower confidence bound}
\acrodefplural{lcb}[LCBs]{lower confidence bounds}
\acro{kl}[KL]{Kullback-Leibler}
\acro{cmab}[CMAB]{combinatorial \ac{mab}}
\acrodefplural{cmab}[CMABs]{combinatorial \acp{mab}}
\end{acronym}

\maketitle

\begin{abstract}
We consider the distributed \acl{sgd} problem, where a {\master} distributes gradient calculations among $\workercnt$ workers. %
By assigning tasks to all the workers and waiting only for the $\batchcnt$ fastest ones, the {\master} can trade-off the algorithm's error with its runtime by gradually increasing $\batchcnt$ as the algorithm evolves. However, this strategy, referred to as \emph{adaptive $k$-sync}, neglects the cost of unused computations and of communicating models to workers that reveal a straggling behavior. We propose a cost-efficient scheme that assigns tasks only to $\batchcnt$ workers, and gradually increases $\batchcnt$. %
We introduce the use of a combinatorial multi-armed bandit model to learn which workers are the fastest while assigning gradient calculations. %
Assuming workers with exponentially distributed response times parameterized by different means, we give empirical and theoretical guarantees on the regret of our strategy, i.e., the extra time spent to learn the mean response times of the workers. %
Furthermore, we propose and analyze a strategy applicable to a large class of response time distributions. Compared to adaptive $k$-sync, our scheme achieves significantly lower errors with the same computational efforts and less downlink communication while being inferior in terms of speed.
\end{abstract}


\section{Introduction}

We consider a distributed machine learning setting, in which a central entity, referred to as the \textit{{\master}}, possesses a large amount of data on which it wants to run a machine learning algorithm. To speed up the computations, the {\master} distributes the computation tasks to several \textit{worker} machines. The workers compute smaller tasks in parallel and send back their results to the {\master}, which then aggregates the results to obtain the desired result of the large computation. A naive distribution of the tasks to the workers suffers from the presence of stragglers, i.e., slow or even unresponsive workers \cite{Dean2013, Chen2017}.

The negative effect of stragglers can be mitigated by assigning redundant computations to the workers and ignoring the response of the slowest ones, e.g., \cite{Lee2016,tandon2017}.
However, in gradient descent algorithms, assigning redundant tasks to the workers can be avoided when a (good) estimate of the gradient loss function is sufficient. On a high level, gradient descent is an iterative algorithm requiring the {\master} to compute the gradient of a loss function at every iteration based on the current model. Simply ignoring the stragglers is equivalent to \ac{sgd} \cite{robbins1951stochastic,cotter2011better}, which advocates computing an estimate of the gradient of the loss function at every iteration~\cite{Chen2017,Dutta2018}. As a result, \ac{sgd} trades-off the time spent per iteration with the total number of iterations for convergence, or until a desired result is reached.
The authors of \cite{Hanna2020} show that for distributed \ac{sgd} algorithms, it is faster for the {\master} to assign tasks to all the workers, but wait for only a small subset of the workers to return their results. In the strategy proposed in \cite{Hanna2020}, called \textit{adaptive $k$-sync}, in order to improve the convergence speed, the {\master} increases the number of workers it waits for as the algorithm evolves in iterations. Despite reducing the run-time of the algorithm, i.e., the total time needed to reach the desired result, this strategy requires the {\master} to transmit the current model to all available workers and pay for all computational resources while only using the computations of the fastest ones.

In this work, we take into account the cost of employing workers and for transferring the current model to the workers. In contrast to~\cite{Hanna2020}, we propose a communication- and computation-efficient scheme that distributes tasks only to the fastest workers and waits for the completion of all their computations. However, in practice, the {\master} does not know in advance which workers are the fastest. To this end, we introduce the use of a stochastic \ac{mab} framework to learn the speed of the workers while efficiently assigning them computational tasks. Stochastic \acp{mab}, introduced in~\cite{Thompson1933}, are iterative algorithms initially designed to maximize the gain of a user gambling with multiple slot machines, termed ``armed bandits''. At each iteration, the user is allowed to pull one arm from the available set of armed bandits. Each arm pull yields a random reward following a known distribution with an unknown mean. The user wants to design a strategy to learn the expected reward of the arms while maximizing the accumulated rewards. Stochastic \acp{cmab} were introduced in \cite{Anantharam1987} and model the behavior when a user seeks to find a combination of arms that reveals the best overall expected reward.

Following the literature on distributed computing~\cite{Liang2014,Lee2016}, we model the response times of the workers by independent and exponentially distributed random variables. We additionally assume that the workers are heterogeneous, i.e., have different mean response times. To apply \acp{mab} to distributed computing, we model the rewards by the response times and aim to minimize the rewards. Under this model, we show that compared to \textit{adaptive $k$-sync}, using a \ac{mab} to learn the mean response times of the workers on the fly cuts the average cost (reflected by the total number of worker employments) but comes at the expense of significantly increasing the total run-time of the algorithm.

\IEEEpubidadjcol

\subsection{Related Work}

\subsubsection{Distributed Gradient Descent}
Assigning redundant tasks to the workers and running distributed gradient descent is known as gradient coding \cite{tandon2017,ye2018communication, Raviv2020, Ozfatura2020,Amiri2019, Li2018}. Approximate gradient coding is introduced to reduce the required redundancy and run \ac{sgd} in the presence of stragglers \cite{Bitar2020StochasticGC,Maity2019,Charles2017, Wang2019a,Wang2019,Horii2019,Ozfatura2019,Ozfatura2021}. The schemes in \cite{Amiri2019,Li2018} use redundancy but no coding to avoid encoding/decoding overheads. However, assigning redundant computations to the workers increases the computation time spent per worker and may slow down the overall computation process. Thus, \cite{Dutta2018,Chen2017,Hanna2020} advocate running distributed \ac{sgd} without redundant task assignment to the workers. In \cite{Dutta2018}, the convergence speed of the algorithm is analyzed in terms of the wall-clock time rather than the number of iterations. It is assumed that the {\master} waits for $\batchcnt$ out of $\workercnt$ workers and ignores the rest. The authors of \cite{Hanna2020} show that gradually increasing $\batchcnt$, i.e., gradually decreasing the number of tolerated stragglers as the algorithm evolves, increases the convergence speed of the algorithm. In this work, we consider a similar analysis to the one in~\cite{Hanna2020}; however, instead of assigning tasks to all the workers and ignoring the stragglers, we require the {\master} to only employ (assign tasks to) the required amount of workers. To learn the speed of the workers and choose the fastest ones, we use ideas from the literature on \acp{mab}.

\subsubsection{\acp{mab}}
Since their introduction in \cite{Thompson1933}, \acp{mab} have been extensively studied for decision-making under uncertainty. A \ac{mab} strategy is evaluated by its \emph{regret} defined as the difference between the actual cumulative reward and the one that could be achieved should the user know the expected reward of the arms a priori. The works of \cite{Lai1985,Agrawal1995} introduced the use of \acp{ucb} based on previous rewards to decide which arm to pull at each iteration. Those schemes are said to be asymptotically optimal since their regret becomes negligible as the number of iterations goes to infinity. In \cite{Auer2002}, the regret of a \ac{ucb} algorithm is bounded for a finite number of iterations. Subsequent research aims to improve on this by introducing variants of \acp{ucb}, e.g., %
\acs{kl}-\ac{ucb} \cite{Garivier2013, Garivier2013ext} which is based on \ac{kl}-divergence. 
While most of the works assume a finite support for the reward, %
\acp{mab} with unbounded rewards were studied in \cite{Garivier2013, Garivier2013ext,  Jouini2012, Bubeck2012}, where in the latter the variance factor is assumed to be known.
In the class of \acp{cmab}, the user is allowed to pull multiple arms with different characteristics at each iteration. The authors of \cite{Anantharam1987} extended the asymptotically efficient allocation rules of \cite{Lai1985} to a \ac{cmab} scenario.
General frameworks for the \ac{cmab} with bounded reward functions are investigated in \cite{Chen2016, Kveton2015, Chen2013, Chen2018}. The analysis in \cite{Gai2010a, Gai2012} for linear reward functions with finite support is an extension of the classical \ac{ucb} strategy, and comes closest to our work. %

 \subsection{Contributions and Outline}
After a description of the system model in \cref{sec:model_preliminaries}, we introduce in \cref{sec:policy_results} a \ac{cmab} model based on \acp{lcb} to reduce the cost of distributed gradient descent, measured in terms of the number of worker employments, whether the results of the corresponding computations carried out by the workers are used by the main node or not. Our cost-efficient policy increases the number of employed workers as the algorithm evolves. In \cref{sec:policy}, we introduce and theoretically analyze an \ac{lcb} that is particularly suited to exponential distributions and easy to compute for the master. To improve the performance of our \ac{cmab}, we investigate in \cref{sec:improved_policy} an \ac{lcb} that is based on \ac{kl}-divergence, and generalizes to all bounded reward distributions and those belonging to the canonical exponential family. This comes at the expense of a higher computational complexity for the main node. In \cref{sec:simulations}, we provide simulation results for linear regression to underline our theoretical findings. \cref{sec:conclusion} concludes the paper.

\section{System Model and Preliminaries} \label{sec:model_preliminaries}

\textbf{Notations.} Vectors and matrices are denoted in bold lower and upper case letters, e.g., $\mathbf{z}$ and $\mathbf{Z}$, respectively. For integers $\kappa$, $\tau$ with $\kappa < \tau$, the set $\curly{\kappa, \kappa+1, \dots , \tau}$ is denoted by $[\kappa, \tau]$, and $\sq{\tau} \define \curly{1, \dots, \tau}$. Sub-gamma distributions are expressed by shape $\alpha$ and rate $\beta$, i.e., $\subgamma{\alpha}{\beta}$, and sub-Gaussian distributions by variance $\sigma^2$, i.e., $\subgaussian{\sigma^2}$. The identity function $\identity{z}$ is $1$ if $z$ is true, and $0$ otherwise. Throughout the paper, we use the terms arm and worker interchangeably. %

We denote by $\samplematrix \in \mathbb{R}^{\samplecnt \times \samplesize}$ a data matrix  with $\samplecnt$ samples, where each sample $\samplevec \in \mathbb{R}^{\samplesize}$, $\sampleidx \in \sq{1, \samplecnt}$, is the $\ell$-th row of $\mathbf{X}$ and by $\samplelabelvec \in \mathbb{R}^\samplecnt$ the vector containing the labels $\samplelabel$ for every sample $\samplevec$. The goal is to find a model $\model \in \mathbb{R}^\samplesize$ that minimizes an additively separable loss function $\lossfct \define \sum_{\sampleidx = 1}^{\samplecnt} F(\samplevec, \samplelabel, \model)$, i.e., to find $\modelopt = \argmin_{\model \in \mathbb{R}^\samplesize} \lossfct$.

To enable flexible distributed computing schemes that use at most $\budget$ workers\footnote{For ease of analysis, we assume that $\budget$ divides $\samplecnt$. This can be satisfied by adding all-zero rows to $\samplematrix$ and corresponding zero labels to $\samplelabelvec$.} out of $\workercnt$ available in parallel, we employ mini-batch gradient descent.
At iteration $\iter$, the {\master} employs a set of workers, indexed by $\iterworkers$, $\card{\iterworkers} \leq \budget$. Every worker $\armidx \in \iterworkers$ computes a partial gradient using a random subset (batch) of $\samplematrix$ and $\samplelabelvec$ consisting of $\batchsize = \frac{\samplecnt}{\budget}$ samples. The data $\samplematrix$ and $\samplelabelvec$ is stored on a shared memory, and can be accessed by all $\workercnt$ workers. Worker $\armidx \in \iterworkers$ computes a gradient estimate $\nabla \workerlossfct[\iter]$ based on subset $\workermatrix \in \mathbb{R}^{\batchsize \times \samplesize}$ of $\samplematrix$, subset $\workervec \in \mathbb{R}^{\batchsize}$ of $\samplelabelvec$ and the model $\model_{\iter}$ at iteration $\iter$. The {\master} waits for $\nonstragglers \subseteq \iterworkers$ responsive workers and updates the model $\model$ as
\begin{align}
    \model_{\iter+1} &= \model_{\iter} - \frac{\lr}{|\mathcal{R}(j)| \cdot \batchsize} \sum\limits_{\armidx \in \mathcal{R}(j)} \nabla \workerlossfct[\iter] \label{eq:update_rule} \\
    &= \model_{\iter} - \frac{\lr}{\card{\nonstragglers} \cdot \batchsize} \sum\limits_{\armidx \in \nonstragglers} \sum\limits_{\sampleidx \in \mathcal{V}_{\armidx, \iter}} \nabla \lossfctsample, \nonumber
\end{align} 
where $\lr$ denotes the learning rate and by $\mathcal{V}_{\armidx, \iter}$ we denote the set of indices of all samples in $\workermatrix$. According to \cite{Bottou2018, Dutta2018}, fixing the value of $\card{\nonstragglers} = \batchcnt$ and running $\iter$ iterations of gradient descent with a mini-batch size of $\batchsize \cdot \batchcnt$ results in an expected deviation from the optimal loss $F^\star$ bounded as\footnote{This holds under the assumptions detailed in \cite{Bottou2018, Dutta2018}, i.e., a Lipschitz-continuous gradient with bounds on the first and second moments of the objective function characterized by $\lipschitz$ and $\gradvar$, respectively, strong convexity with parameter $\convexity$, the stochastic gradient being an unbiased estimate, and a sufficiently small learning rate $\lr$.}
\begin{align}
    \!\!\expdev{\batchcnt}{\iter} &= \mathbb{E}[F(\batchcnt, \model_\iter) - F^\star] \nonumber \\
    &\leq \!\! \underbrace{\frac{\eta \lipschitz \gradvar}{2 \convexity \batchcnt \batchsize}}_{\text{error floor}} \!\! + \underbrace{(1-\eta \convexity)^{\iter} \left(F(\initvec) - F^\star - \frac{\eta \lipschitz \gradvar}{2 \convexity \batchcnt \batchsize} \right)}_{\text{transient behavior}}. \label{eq:convergence}
\end{align}
As the number of iterations goes to infinity, the influence of the transient behavior vanishes and what remains is the contribution of the error floor.

\section{CMAB for Distributed Learning} \label{sec:policy_results}
We group the iterations into $\budget$ rounds, such that at iterations within round  $\roundidx \in \sq{1, \budget}$, the {\master} employs $\card{\iterworkers} = \roundidx$ workers and waits for all of them to respond, i.e., $\iterworkers = \nonstragglers$. 
As in \cite{Hanna2020}, we let each round $\roundidx$ run for a predetermined number of iterations. That is, at a switching iteration $\iter=\timehorizonround$, the algorithm advances to round $\roundidx+1$. We define $\timehorizon_0 \define 0$, i.e., the algorithm starts in round one, and $\timehorizon_\budget$ as the last iteration, i.e., the algorithm ends in round $\budget$. %
The total budget $\totalbudget$ is defined as $\totalbudget \define \sum_{\roundidx = 1}^\budget \roundidx \cdot \br{\timehorizonround[\roundidx] - \timehorizonround[\roundidx - 1]}$, which gives the total number of worker employments.%

We assume exponentially distributed response times of the workers; that is, the response time $\armresp$ of worker $\armidx$ in iteration $\iter$, resulting from the sum of communication and computation delays, follows an exponential distribution with rate $\lambda_\armidx$ and mean $\mui= \frac{1}{\lambda_\armidx}$, i.e., $\armresp \sim \exp(\lambda_\armidx)$. The minimum rate of all the workers is $\lambda_{\min} \define \min_{\armidx \in \sq{1, \workercnt}} \lambda_\armidx$. The goal is to assign tasks only to the $\roundidx$ fastest workers. We denote by policy $\policy$ a decision process that chooses the $\roundidx$ expected fastest workers. The optimal policy $\policyopt$ assumes knowledge of the $\mui$'s and chooses $\roundidx$ workers with the smallest $\mui$'s. However, in practice the $\mui$'s are unknown in the beginning. Thus, our objective is two-fold. First, we want to find confident estimates $\muihat$ of the mean response times $\mui$ to correctly identify (explore) the fastest workers, and second, we want to leverage (exploit) this knowledge to employ the fastest workers as much as possible, rather than investing in unresponsive/straggling workers. To trade-off this exploration-exploitation dilemma, we utilize the \ac{mab} framework where each arm $\armidx \in [1,\workercnt]$ corresponds to a different worker and $\roundidx$ arms are pulled at each iteration. %
A \emph{superarm} $\superarm \subseteq \sq{\workercnt}$ with $\card{\superarm} = \roundidx$ is the set of indices of the arms pulled at iteration $\timeidx$, and $\superarmbest$ is the optimal choice containing the indices of the $\roundidx$ workers with the smallest means. For every worker, we maintain a counter $\pullcounter$ for the number of times this worker has been employed until iteration $\timeidx$, i.e., $\pullcounter= \sum_{\tmpvar = 1}^\iter \identity{\armidx \in \superarmvar^\roundidx (\tmpvar)}$, and a counter $\respsum$ for the sum of its response times, i.e., $\respsum = \sum_{\tmpvar = 1}^\iter \identity{\armidx \in \superarmvar^\roundidx (\tmpvar)} \armresp[\tmpvar]$. The \ac{lcb} of a worker is a measure based on the empirical mean $\muihat(\iter) = \frac{\respsum}{\pullcounter}$ and the number of samples $\pullcounter$ chosen such that the true mean $\mui$ is unlikely being smaller. As the number of samples grows, the \ac{lcb} of worker $\armidx$ approaches $\muihat$. A policy $\policy$ is a rule to compute and update the \acp{lcb} of the $\workercnt$ workers such that at iteration $\timeidx \in [\timehorizonround[\roundidx - 1]+1, \timehorizonround[\roundidx]]$ the $\roundidx$ workers with the smallest \acp{lcb} are pulled. %
The choice of the confidence bounds significantly affects the performance of the model and will be analyzed in \cref{sec:policy,sec:improved_policy}. A summary of the \ac{cmab} policy and the steps executed by the workers is given in \cref{alg:combinatorial_strategy}.%

\ifdouble
\begin{algorithm}[!htb]
\caption{Combinatorial Multi-Armed Bandit Policy}\label{alg:combinatorial_strategy}
\begin{algorithmic}[1]
    \Require Number of workers $\workercnt$, budget $\budget \leq \workercnt$
    \Initialize $\forall \; \armidx \in [\workercnt]$: $\cbshort[0]{\armidx} \gets -\infty$ and $\pullcountervar[0]{\armidx} \gets 0$
    \For{$\roundidx = 1, \dots, \budget$} \Comment{\parbox[t]{.55\linewidth}{Run (combinatorial) \ac{mab} with $\workercnt$ arms while pulling $\roundidx$ at a time}\vspace{1.2mm}}
    \For{$\timeidx = \timehorizonround[\roundidx-1]\!+\!1, \dots, \timehorizonround$}
    \State $\forall \armidx \in \sq{\workercnt}:$ calculate $\cbshort[\timeidx-1]{\armidx}$ %
    \State \parbox[t]{\dimexpr.9\linewidth-\algorithmicindent\relax}{Choose $\superarm$, i.e., $\roundidx$ workers with the minimum $\cbshort[\timeidx-1]{\armidx}$ where $\armidx \in \sq{\workercnt}$}\strut
    \State \parbox[t]{\dimexpr.9\linewidth-\algorithmicindent\relax}{Every worker $\armidx \in \superarm$ computes a gradient estimate $\nabla \workerlossfct[\iter]$ and sends \vspace{0mm}\\\vspace{0mm}it to the main node}\strut
    \State $\forall \armidx \in \superarm$: Observe response time $\armresp$
    \State \parbox[t]{\dimexpr.9\linewidth-\algorithmicindent\relax}{$\forall \armidx \in \superarm$: Update statistics, i.e., $\pullcounter = \pullcountervar[\timeidx-1]{\armidx} + 1$, $\muihat(\timeidx) = \frac{\armresp + \pullcountervar[\timeidx-1]{\armidx} \cdot \muihat(\timeidx-1)}{\pullcounter}$}\strut
    \State Update model $\model_\iter$ according to \eqref{eq:update_rule}
    \EndFor
    \EndFor
\end{algorithmic}
\end{algorithm}
\else
\begin{algorithm}[!htb]
\caption{Combinatorial Multi-Armed Bandit Policy}\label{alg:combinatorial_strategy}
\begin{algorithmic}[1]
    \Require Number of workers $\workercnt$, budget $\budget \leq \workercnt$
    \Initialize $\forall \; \armidx \in [\workercnt]$: $\cbshort[0]{\armidx} \gets -\infty$ and $\pullcountervar[0]{\armidx} \gets 0$
    \For{$\roundidx = 1, \dots, \budget$} \Comment{\parbox[t]{.7\linewidth}{Run (combinatorial) \ac{mab} with $\workercnt$ arms while pulling $\roundidx$ at a time}\vspace{0.0mm}}
    \For{$\timeidx = \timehorizonround[\roundidx-1]\!+\!1, \dots, \timehorizonround$}
    \State $\forall \armidx \in \sq{\workercnt}:$ calculate $\cbshort[\timeidx-1]{\armidx}$ %
    \State \parbox[t]{\dimexpr\linewidth-\algorithmicindent\relax}{Choose $\superarm$, i.e., $\roundidx$ workers with the minimum $\cbshort[\timeidx-1]{\armidx}$ where $\armidx \in \sq{\workercnt}$}\strut
    \State \parbox[t]{\dimexpr\linewidth-\algorithmicindent\relax}{Every worker $\armidx \in \superarm$ computes a gradient estimate $\nabla \workerlossfct[\iter]$ and sends \vspace{-2mm}\\\vspace{2.5mm}it to the main node}\strut
    \State $\forall \armidx \in \superarm$: Observe response time $\armresp$
    \State \parbox[t]{\dimexpr\linewidth-\algorithmicindent\relax}{$\forall \armidx \in \superarm$: Update statistics $\pullcounter = \pullcountervar[\timeidx-1]{\armidx} + 1$, $\muihat(\timeidx) = \frac{\armresp + \pullcountervar[\timeidx-1]{\armidx} \cdot \muihat(\timeidx-1)}{\pullcounter}$}\strut
    \State Update model $\model_\iter$ according to \eqref{eq:update_rule}
    \EndFor
    \EndFor
\end{algorithmic}
\end{algorithm}
\fi

In contrast to most works on \acp{mab}, we minimize an unbounded objective, i.e., the overall computation time $\superarmresp \define \max_{\armidx \in \superarm} \armresp$ at iteration $\iter$. This corresponds to waiting for the slowest worker. The expected response time of a superarm $\superarm$ is then defined as $\mu_{\superarm} \define \mathbb{E}[\superarmresp]$ and can be calculated according to \cref{prop:max_mean}.

\begin{proposition} \label{prop:max_mean} The mean of the maximum of independently distributed exponential random variables with different means, indexed by a set $\tmpsetsecond$, i.e., $\rvtmp_\tmpvarthird \sim \exp\br{\lambda_\tmpvarthird}$, $\tmpvarthird \in \tmpsetsecond$, is given as
\begin{align}
\label{eq:max_exp}
   \E{\max\limits_{\tmpvarthird \in \tmpsetsecond} \rvtmp_\tmpvar} = \sum\limits_{\tmpsetthird \in \ps{\tmpsetsecond}\setminus\emptyset} \br{-1}^{\vert \tmpsetthird \vert - 1} \frac{1}{\sum_{\tmpvarforth \in \tmpsetthird} \lambda_\tmpvarforth},
\end{align}
with $\ps{\tmpsetsecond}$ denoting the power set of $\tmpsetsecond$.
\begin{proof}
The proof is given in \cref{sec:proof_prop_max_mean}.
\end{proof}
\end{proposition}

\begin{proposition} \label{prop:variance} 
The variance of the maximum of independently distributed exponential random variables with different means, indexed by a set $\tmpsetsecond$, i.e., $\rvtmp_\tmpvarthird \sim \exp\br{\lambda_\tmpvarthird}$, $\tmpvarthird \in \tmpsetsecond$, is given as
\begin{align*}
    \Var{\max\limits_{\tmpvarthird \in \tmpsetsecond} \rvtmp_\armidx} = \!\!\!\!\!\! \sum\limits_{\tmpsetthird \in \ps{\tmpsetsecond}\setminus \emptyset} \!\!\!\!\!\! \br{-1}^{\vert \tmpsetthird \vert - 1} \! \br{\!2\! \br{\frac{1}{\sum_{\tmpvarforth \in \tmpsetthird} \lambda_\tmpvarforth}}^2 \!\!\! - \frac{1}{\sum_{\tmpvarforth \in \tmpsetthird} \lambda_\tmpvarforth}}
\end{align*} with $\ps{\tmpsetsecond}$ denoting the power set of $\tmpsetsecond$.
\begin{proof}
The proof follows similar lines as for \cref{prop:max_mean} and is omitted for brevity.
\end{proof}
\end{proposition}

The suboptimality gap of a chosen (super-)arm describes the expected difference in time compared to the optimal choice.
\begin{definition} \label{def:gaps}
For a superarm $\superarm$ and for $\superarmvar^\roundidx_{\text{worst}}$ defined as the set of indices of the $\roundidx$ slowest workers, we define the following superarm suboptimality gaps
\begin{align}
    \supergap &\define \supergaplong, \nonumber \\
    \supergapmax &\define \mu_{\superarmvar^\roundidx_{\text{worst}}} - \musbest. \label{eq:supergapmax}
\end{align}
For $\element\leq \roundidx$, $\superarmelement$ and $\superarmbestelement$ denote the indices of the $\element^\text{th}$ fastest worker in $\superarm$ and $\superarmbest$, respectively. Then, we define the suboptimality gap for the \emph{employed} arms as %
\begin{align*}
    \delta_{\superarmelement} &\define \mu_{\superarmelement} - \mu_{\superarmbestelement}.
\end{align*}
Let $\superarmspace$ denote the set of all superarms with cardinality $\roundidx$. %
We define the minimum suboptimality gap for \emph{all} the arms as
\begin{align}
    \delta_{\min} &\define \min\limits_{\roundidx \in [\budget], \anysuperarm \in \superarmspace} \ \  \min\limits_{ \element \in [\roundidx] : \mu_{\anysuperarm_\element} > \mu_{\superarmbestelement}} \delta_{\anysuperarm_\element}. \label{eq:deltamin}
\end{align}
\end{definition}

\begin{example}
For mean worker computation times given by $\tmpset = \curly{\mu_1, \dots, \mu_\workercnt}$,
we obtain $
    \delta_{\min} \geq \min_{\tmpa,\tmpb \in \tmpset: \tmpa > \tmpb} \tmpa - \tmpb.
$
\end{example}

\begin{definition}\label{def:regret}
We define the regret $\regret$ of a policy $\policy$ run until iteration $\iter$ as the expected difference in run-time of the policy $\policy$ compared to the optimal policy $\policyopt$, i.e.,
\begin{equation*}
    \label{eq:def_regret_cmab}
    \regret \!=\! \E{\sum_{\tmpvar=1}^\timeidx \respvar_{\anysuperarm(\tmpvar)}^\tmpvar} - \!\!\!\!\!\!\!\!\!\!\!\! \smashoperator[r]{\sum\limits_{\,\,\,\,\,\,\,\roundidx \in [\budget]: \timeidx > \timehorizonround[\roundidx-1]}} \br{\min\curly{\timeidx, \timehorizonround}-\timehorizonround[\roundidx-1]} \!\! \min\limits_{\anysuperarm \in \superarmspace} \! \mu_{\anysuperarm}.
\end{equation*}
\end{definition}

\cref{def:regret} quantifies the overhead in total time spent by $\policy$ to learn the average speeds of the workers and will be analyzed in \cref{sec:policy,sec:improved_policy} for two different policies, i.e., choices of \acp{lcb}. We provide in \cref{theorem:time_bound} a run-time guarantee of an algorithm using a \ac{cmab} for distributed learning as a function of the regret $\regret$ and the number of iterations $\iter$.

\begin{theorem} \label{theorem:time_bound}
Given a desired $\epsilon>0$, the time until policy $\policy$ reaches iteration $\iter$ is bounded from above as
\begin{equation*}
  \timevar_\iter^\policy \! \leq \! R_\iter^{\policy} +\!\! \sum\limits_{\roundidx = 1}^\budget \! \identity{\iter > \timehorizonround[\roundidx-1]} \mu_{\superarmbest} \!\!\br{\min\curly{\iter, \timehorizonround}-\timehorizonround[\roundidx-1]} \!\br{1+\epsilon}
\end{equation*}
with probability $$\Pr(\iter) \geq \prod\limits_{r \in \sq{\budget}: \iter > \timehorizonround[\roundidx-1]} \br{1 - \frac{\sigma_{\superarmbest}^2}{\mu_{\superarmbest}^2 \br{\min\curly{\iter, \timehorizonround}-\timehorizonround[\roundidx-1]} \epsilon^2}}.$$ The mean $\mu_{\superarmbest}$ and variance $\sigma_{\superarmbest}^2$ can be calculated according to \cref{prop:max_mean,prop:variance}.
\begin{proof}
The proof is given in \cref{sec:proof_time_bound}.
\end{proof}
\end{theorem}

To give a complete performance analysis, we provide in \cref{theorem:error_bound} a handle on the expected deviation from the optimal loss as a function of number of iterations $\iter$. Combining the results of \cref{theorem:time_bound,theorem:error_bound}, we obtain a measure on the expected deviation from the optimal loss with respect to time.

\begin{remark} \label{theorem:error_bound}
The expected deviation from the optimal loss at iteration $\iter$ in round $\roundidx$ of an algorithm using \ac{cmab} for distributed learning can be bounded by  $\expdev{\batchcnt}{\iter^\prime}$ as in \eqref{eq:convergence} for $\batchcnt=\roundidx$ and $\iter^\prime = \iter_{\roundidx,\text{begin}} + \iter - \timehorizonround[\roundidx-1]$, where $\iter_{1, \text{begin}}=0$, and for $\roundidx \in \sq{2, \budget}$,
\begin{align*}
    \iter_{\roundidx, \text{begin}} = &\frac{\log\br{\frac{\eta \lipschitz \gradvar}{2 \convexity \batchsize}\br{\frac{1}{\roundidx-1} - \frac{1}{\roundidx}} + \alpha^{\iter_{\roundidx-1,\text{end}}} \br{E_0 - \frac{\lr \lipschitz \gradvar}{2 \convexity (\roundidx-1) \batchsize}}}}{\log \br{\alpha}}
    \\
    &-\frac{\log\br{E_0 - \frac{\lr \lipschitz \gradvar}{2 \convexity \roundidx \batchsize}}}{\log \br{\alpha}},
\end{align*}
with $\iter_{\roundidx, \text{end}} \define \iter_{\roundidx, \text{begin}} + \timehorizonround[\roundidx] - \timehorizonround[\roundidx-1]$, $E_0 \define F(\initvec) - F^\star$ and $\alpha \define 1-\lr \convexity$. This is because at each round $\roundidx$, the algorithm follows the convergence behavior of an algorithm with mini-batch of fixed size $\roundidx \batchsize$. For algorithms with fixed mini-batch size, we only need the number of iterations ran in order to bound the expected deviation from the optimal loss. However, in round $\roundidx>1$ and iteration $\iter$, the algorithm has advanced differently than with a constant mini-batch of size $\roundidx \batchsize$. Thus, we need to recursively compute the equivalent number of iterations $\iter^\prime \in \sq{\timehorizonround[\roundidx-1]+1, \timehorizonround}$ that have to be run for a fixed mini-batch of size $\roundidx \batchsize$ to finally apply~\eqref{eq:convergence}. Therefore, we have to compute $\iter_{\roundidx,\text{begin}}$, which denotes the iteration for a fixed batch size $\roundidx \batchsize$ with the same error as for a batch size of $(\roundidx-1) \batchsize$ at the end of the previous round $\roundidx-1$, denoted by $\iter_{\roundidx-1,\text{end}} \define \iter_{\roundidx-1, \text{begin}} + \timehorizonround[\roundidx-1] - \timehorizonround[\roundidx-2]$. To calculate $\iter_{\roundidx,\text{begin}}$ it has to hold that $\expdev{\roundidx}{\iter_{\roundidx, \text{begin}}} = \expdev{\roundidx-1}{\iter_{\roundidx-1,\text{end}}}$. This can be repeated recursively, until we can use $\iter_{1,\text{end}}$. For round $\roundidx=1$, the problem is trivial.\footnote{Alternatively, one could also use the derivation in \cite[Equation 4.15]{Bottou2018} and recursively bound the expected deviation from the optimal loss in round $\roundidx$ based on the expected deviation at the end of the previous round $\roundidx-1$, i.e, use $\mathbb{E}[F(\model_{\timehorizonround[\roundidx-1]}) - F^\star]$ instead of $E_0$.}

\end{remark}

In this section, the \acp{lcb} were treated as a black box. In \cref{sec:policy,sec:improved_policy}, we present two different \ac{lcb} policies together with respective performance guarantees.

\begin{remark}
The explained policies can be seen as an \ac{sgd} algorithm which gradually increases the mini-batch size. In the machine learning literature, this is one of the approaches considered to optimize convergence. Alternatively, one could also use $\budget$ workers with a larger learning rate from the start and gradually decrease the learning rate to trade-off the error-floor in \eqref{eq:convergence} with run-time. For the variable learning rate approach, one can use a slightly adapted version of our policies, where $\card{\superarm} = \budget$ is fixed. In case the goal is to reach a particular error floor, our simulations show that the latter approach reaches this error faster than the former. This, however, only holds under the assumption that the chosen learning rate in \eqref{eq:update_rule} is sufficiently small, i.e., the scaled learning rate at the beginning of the algorithm still leads to convergence. However, if one seeks to optimize the convergence speed at the expense of reaching a slightly higher error floor, simulations show that decaying the learning rate is slower because the learning rate is limited to ensure convergence. Optimally, one would combine both approaches by starting with the maximum possible learning rate, gradually increasing the number of workers per iteration until reaching $\budget$, and then decreasing the learning rate to reach the best possible error floor. %
\end{remark}

\section{Confidence Radius Based Policy}\label{sec:policy}

\renewcommand{\cbradcomb}[1]{\ensuremath{\sqrt{\frac{4 \cbscale}{#1}} + \frac{2 \cbscale}{#1}}}

Motivated by \cite{Auer2002}, we present a confidence radius based policy $\cmabpolicy$ that is computationally light for the main node.
With this policy, in iteration $\timeidx \in [\timehorizonround[\roundidx - 1]+1, \timehorizonround[\roundidx]]$ the superarm $\superarm$ is chosen as the $\roundidx$ arms with the lowest \acp{lcb} calculated as
\begin{equation*}
    \cbshort[\timeidx-1]{\armidx} \define 
    \begin{cases}
    -\infty & \text{if } \pullcountervar[\timeidx-1]{\armidx}=0 \\
    \cblong & \text{otherwise},
    \end{cases}
\end{equation*}
where $\muihat(\timeidx) \define \frac{\respsum[\timeidx]}{\pullcountervar[\timeidx]{\armidx}}$. The choice of the confidence radius $\cbradshort{\armidx}{\timeidx}$ affects the performance of the policy and is, based on the underlying reward distribution\footnote{By this particular choice, we can prove a bounded regret in the setting of minimizing outcomes subject to an exponential distributions.}, chosen as $\cbradshort{\armidx}{\timeidx} \define \cbradcomb{\pullcountervar{\armidx}}$, with $\cbscale = 2\log(\timeidx)$. 
The estimates $\muihat(\iter)$ and the confidence radii are updated after every iteration according to the responses of the chosen workers. %
We give a performance guarantee for this confidence bound choice in terms of the regret in~\cref{theorem:regret_bound}.

\newcommand{\roundoffset}{\frac{48\log(\timeidx)}{\min\{\delta_{\min}^2, \delta_{\min}\}}}
\newcommand{\roundoffsetbounded}{\frac{48\log(\min\curly{\timeidx, \timehorizonround})}{\min\curly{\delta_{\min}^2, \delta_{\min}}}}
  
\begin{theorem} \label{theorem:regret_bound}
The regret of the \ac{cmab} policy $\cmabpolicy$ with gradually increasing superarm size and arms chosen based on \acp{lcb} with radius $\cbradshort{\armidx}{\iter} \define \cbradcomb{\pullcountervar{\armidx}}$ where $\cbscale = 2\log(\timeidx)$, and assuming\footnote{The assumption $\lambda_{\min} \geq 1$ is needed for our proof to hold. In practice, this assumption amounts to choosing the time unit of our theoretical model such that the average response time of each worker is less than one time unit.} $\lambda_{\min} \geq 1$, is bounded from above as
\ifdouble\begin{align}
    R_\iter^{\cmabpolicy} \leq &\max\limits_{\roundidx \in \sq{\budget}: \timeidx>\timehorizon_{\roundidx-1}} \Delta_{\superarmvar^{\roundidx}, \max} \cdot \workercnt \cdot \vphantom{} \nonumber \\ & \br{\roundoffset + 1 + \tmpvarsecond \cdot \frac{\pi^2}{3}},
\end{align}\else
\begin{align}
    R_\iter^{\cmabpolicy} \leq &\max\limits_{\roundidx \in \sq{\budget}: \timeidx>\timehorizon_{\roundidx-1}} \Delta_{\superarmvar^{\roundidx}, \max} \cdot \workercnt \cdot \vphantom{} \nonumber  \br{\roundoffset + 1 + \tmpvarsecond \cdot \frac{\pi^2}{3}},
\end{align}\fi
where $\tmpvarsecond \define \max_{\roundidx \in \sq{\budget}: \timeidx>\timehorizon_{\roundidx-1}} \roundidx$. %
\begin{proof}
The proof is given in \cref{sec:proof}.
\end{proof}
\end{theorem}

\section{KL-Based Policy} \label{sec:improved_policy}

\newcommand{\cbkl}{\ensuremath{\min\curly{q < \muihat: \pullcountervar[\timeidx]{\armidx} \cdot \kl{p_{\muihat}}{p_{q}} \leq f(\timeidx)}}}

The authors of \cite{Garivier2013} propose to use a \ac{kl}-divergence-based confidence bound for \acp{mab} to improve the regret compared to classical \ac{ucb}-based algorithms. Due to the use of \ac{kl}-divergence, this scheme is applicable to reward distributions that have bounded support or belong to the canonical exponential family. %
Motivated by this, we extend this model to a \ac{cmab} for distributed machine learning and define a policy $\klpolicy$ that calculates \acp{lcb} according to
\begin{align} \label{eq:kl_cb}
    \!\!\!\cbshort{\armidx} \define \cbkl,\!
\end{align}
where $f(\timeidx) = \log(\timeidx) + 3\log\br{\log\br{\timeidx}}$. This confidence bound, i.e., the minimum value for $q$, can be calculated using the Newton procedure for root finding by solving $e(\timeidx, \muihat, q) \define \pullcountervar{\armidx} \cdot \kl{p_{\muihat}}{p_{q}} - \log(\timeidx) = 0$, and is thus computationally heavy for the main node. 
For exponential distributions with probability density function $p$ parametrized by means $\muihat$ and $q$, respectively, the \ac{kl}-divergence is given by $\kl{p_{\muihat}}{p_{q}} = \muihat/q - \log\br{\muihat/q} - 1$. Its derivative can be calculated as $\partial \kl{p_{\muihat}}{p_{q}}/\partial q = 1/q - \muihat/q^2$. With this at hand, the $\tmpvarforth^\text{th}$ Newton update denotes as\footnote{Note that $q_0$ must not be equal to $\muihat$. In case $q_0 = \muihat$, the first update step would be undefined since $\partial \kl{p_{\muihat}}{p_{q_0}}/\partial q_0$ would be $0$. In addition, $q_0$ should be chosen smaller than $\muihat$, e.g., $q_0 = 0.01 \cdot \muihat$.}
\begin{equation*}
    q_{\tmpvarforth} = q_{\tmpvarforth-1} - \frac{\kl{p_{\muihat}}{p_{q_{\tmpvarforth-1}}}}{\partial \kl{p_{\muihat}}{p_{q_{\tmpvarforth-1}}}/\partial q_{\tmpvarforth-1}}.
\end{equation*}

For this policy $\klpolicy$, we give a worst case regret in \cref{theorem:kl_regret}. To ease the notation, we write $\klabbrev{\tmpa}{\tmpb} \define \kl{p_\tmpa}{p_\tmpb}$.

\newcommand{\klspaceleft}{\!\!\!\!\!}
\newcommand{\klspaceright}{\!\!\!\!\!\!\!\!}

\begin{theorem} \label{theorem:kl_regret} Let the response times of the workers be sampled from a finitely supported distribution or a distribution belonging to the canonical exponential family. Then, the regret of the \ac{cmab} policy $\klpolicy$ with gradually increasing superarm size and arms chosen based on a \ac{kl}-based confidence bound $\cbshort{\armidx} \define \cbkl$ with $f(\timeidx) = \log(\timeidx) + 3\log\br{\log\br{\timeidx}}$, $\iter>3$, and $\tmpvarsecond = \!\!\!\! \max\limits_{\roundidx \in \sq{\budget}: \timeidx>\timehorizon_{\roundidx-1}} \!\!\!\! \roundidx$ can be upper bounded as
\ifdouble\begin{align}
    \label{eq:kl_regret_tight}
    \regret &\leq \max\limits_{\roundidx \in \sq{\tmpvarsecond}} \Delta_{\superarmvar^{\roundidx}, \max} \cdot \armcnt \cdot \tmpvarsecond \cdot \left(\vphantom{\frac{\exp\br{-\klepsmin \br{\frac{1 + \epsilon}{\klmax} \br{f(\timeidx) - 1}}}}{1 - \exp(-\klmin/(1 + \epsilon))}}7 \log(\log(\timeidx)) + \frac{1 + \epsilon}{\klmin} f(\timeidx) \right. \nonumber \\
    &\left.  + \, \frac{\exp\br{-\klepsmin \br{\frac{1 + \epsilon}{\klmax} f(\timeidx) - 1}}}{1 - \exp(-\klepsmin)}\right),
\end{align}\else
\begin{align}
    \label{eq:kl_regret_tight}
    \regret &\leq \max\limits_{\roundidx \in \sq{\tmpvarsecond}} \Delta_{\superarmvar^{\roundidx}, \max} \cdot \armcnt \cdot \tmpvarsecond \cdot \!\!\left(\vphantom{\frac{\exp\br{-\klepsmin \br{\frac{1 + \epsilon}{\klmax} \br{f(\timeidx) - 1}}}}{1 - \exp(-\klmin/(1 + \epsilon))}}\!7 \log(\log(\timeidx)) + \frac{1 + \epsilon}{\klmin} f(\timeidx) \right. \nonumber 
    \left.  + \, \frac{\exp\br{-\klepsmin \br{\frac{1 + \epsilon}{\klmax} f(\timeidx) - 1}}}{1 - \exp(-\klepsmin)}\!\right)\!\!,
\end{align}\fi
where $\epsilon$ is a parameter that can be freely chosen and
\begin{align*}
    \klmax &= \klspaceleft\max\limits_{\roundidx \in \sq{\budget}, \tmpset \in \superarmspace, \element \in \sq{\roundidx}: \mu_{\tmpset_\element} > \mu_{\superarmbestelement}} \klspaceright\klabbrev{\mu_{\tmpset_\element}}{\mu_{\superarmbestelement}}, \\
    \klmin &= \klspaceleft\min\limits_{\roundidx \in \sq{\budget}, \tmpset \in \superarmspace, \element \in \sq{\roundidx}: \mu_{\tmpset_\element} > \mu_{\superarmbestelement}} \klspaceright\klabbrev{\mu_{\tmpset_\element}}{\mu_{\superarmbestelement}}, \\
    \klepsmin &= \klspaceleft\min\limits_{\roundidx \in \sq{\budget}, \tmpset \in \superarmspace, \element \in \sq{\roundidx}: \mu_{\tmpset_\element} > \mu_{\superarmbestelement}} \klspaceright\klabbrev{\phi(\epsilon, \mu_{\tmpset_\element}, \mu_{\superarmbestelement})}{\mu_{\tmpset_\element}},
\end{align*}
with $\mu_{\superarmbestelement} < \phi(\epsilon, \mu_{\superarmbestelement}, \mu_{\tmpset_\element}) < \mu_{\tmpset_\element}$ such that $\kl{\phi(\epsilon, \mu_{\superarmbestelement}, \mu_{\tmpset_\element})}{\mu_{\superarmbestelement}} =\frac{ \kl{\mu_{\tmpset_\element}}{\mu_{\superarmbestelement}}}{(1+\epsilon)}$.
\begin{proof}
The proof is given in \cref{sec:proof_kl}.
\end{proof}
\end{theorem}

\section{Numerical Simulations} \label{sec:simulations}

\subsection{Setting} \label{subsec:setting}
Similarly to \cite{Hanna2020}, we consider $\workercnt = 50$ workers with exponentially distributed response times whose means are chosen uniformly at random from $\curly{0.1, 0.2, \cdots, 0.9}$ such that $\lambda_{\min} \geq 1$. We limit the budget to $\budget=20$ parallel computations. We create $\samplecnt = 2000$ samples $\samplevec$ with $\samplesize = 100$ entries, each drawn uniformly at random from $[1, 10]$ with labels $\samplelabel \sim \mathcal{N}\br{\samplevec^T \model^\prime, 1}$, for some $\model^\prime$ drawn uniformly at random from  $[1, 100]^{\samplesize}$. The model $\model$ is initialized uniformly at random as $\initvec \in [1, 100]^\samplesize$ and optimized subject to the least squares loss function $\lossfct \define \frac{1}{2} \Vert \samplematrix \model - \samplelabelvec \Vert_2^2$ with learning rate $\lr=\num{1e-4}$. %
We assess the performance of the model by the error function $\Vert \samplematrix^+ \samplelabelvec - \model \Vert_2$, where $\samplematrix^+$ denotes the pseudo-inverse of $\samplematrix$, that quantifies the gap with the analytical solution, so that the analysis largely does not depend on the data nor the problem. For all the simulations, we present the results averaged over at least ten rounds. %

\subsection{Switching Points}
The switching points $\timehorizonround$, $\roundidx \in \sq{\budget}$, are the iterations in which we advance from round $\roundidx$ to $\roundidx+1$. In \cite{Hanna2020}, Pflug's method \cite{Pflug1990} is used to determine the $\timehorizonround$'s on the fly. %
However, this method is very sensitive to the learning rate \cite{Chee2018,Pesme2020}, and may result in different $\timehorizonround$'s across different runs. While implicit model updates \cite{Chee2018} or alternative criteria \cite{Pesme2020} can avoid this effect, we fix the switching points to ensure comparability across simulation runs.
We empirically determine $\timehorizonround[1]$ and necessary statistics to calculate $\timehorizonround$ for $r\in \sq{2, \budget}$ using~\eqref{eq:convergence}.

\begin{figure}[t]
    \centering
    \includegraphics[width=\imgscale\linewidth]{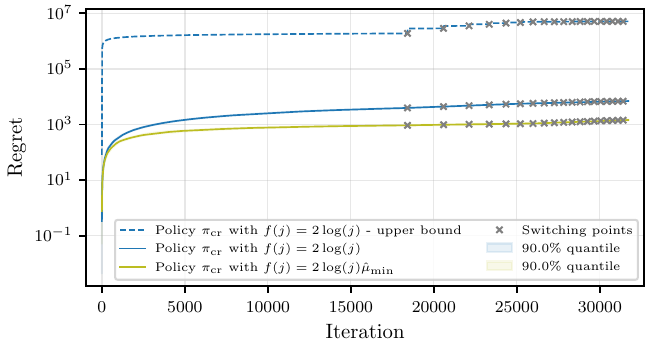}
    \caption{Comparison of the theoretical and simulated regret for $\cmabpolicy$.}
    \label{fig:tail_cb_regret}
    \vspace{-0.3cm}
\end{figure}
\subsection{Simulation Results for Confidence Radius Policy \cmabpolicy} \label{subsec:sim_res}
We first study the \ac{cmab} policy $\cmabpolicy$, which has been introduced in \cref{sec:policy}. Note that the confidence radii are not dependent on the actual mean worker response times. In case the workers have small response times, the confidence radii $\cbradshort{\armidx}{\iter}$ might be very dominant compared to the empirical mean estimates $\muihat$, leading to a frequent employment of suboptimal workers. For practical purposes, it may thus be beneficial to use an adapted confidence radius %
with $\cbscale = 2\log(\timeidx) \hat{\mu}_{\min}$, where $\hat{\mu}_{\min} = \min_{\armidx \in \sq{1, \workercnt}} \muihat$, that balances the confidence radius and the mean estimate.
In \cref{fig:tail_cb_regret}, we compare the theoretical regret guarantee in \cref{theorem:regret_bound} to practical results for both confidence bound choices. As the theoretical guarantee is a worst case analysis, the true performance is underestimated significantly. 
We can see that using $\cbscale = 2\log(\timeidx) \hat{\mu}_{\min}$ significantly improves the regret. However, this comes at the cost of delaying the determination of the fastest workers. While with $\cbscale = 2\log(\iter)$ the policy correctly determined all $\budget$ fastest workers in ten simulation runs, with $\cbscale = 2\log(\iter) \hat{\mu}_{\min}$ in one out of ten simulations the algorithm commits to a worker with a suboptimality gap of $0.1$. This reflects the trade-off between the competing objectives of best arm identification and regret minimization discussed in \cite{Zhong2021}. However, since the fastest workers have been determined eventually with an accuracy of $99.5\%$, the proposed adapted confidence bound seems to be a good choice in practice. Although the theoretical bound is rather loose and deviates from the simulations up to some multiplicative factors, it shows the round-based behavior in a worst-case scenario.

\begin{figure}[t]
    \centering
    \includegraphics[width=\imgscale\linewidth]{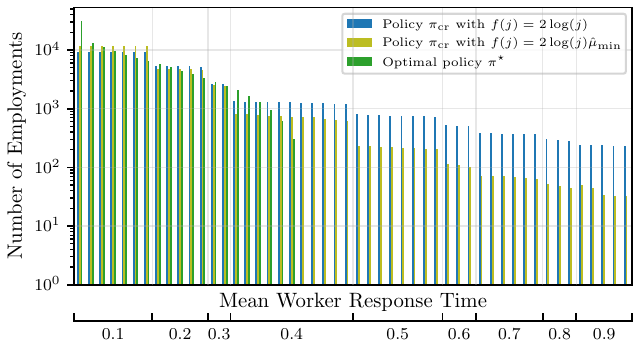}
    \caption{Number of worker employments for $\cmabpolicy$.}
    \label{fig:counters_cmab}
\end{figure}

The number of worker employments is shown in \cref{fig:counters_cmab}, with the workers sorted from fastest to slowest. As expected, compared to the optimal strategy, the \ac{lcb}-based algorithms have to explore all workers including suboptimal ones. With the adapted confidence bound and compared to $\cbscale=2\log(j)$, suboptimal workers are employed significantly less due to the reason above. This also explains the different regrets in \cref{fig:tail_cb_regret}.

\begin{figure}[t]
    \centering
    \includegraphics[width=\imgscale\linewidth]{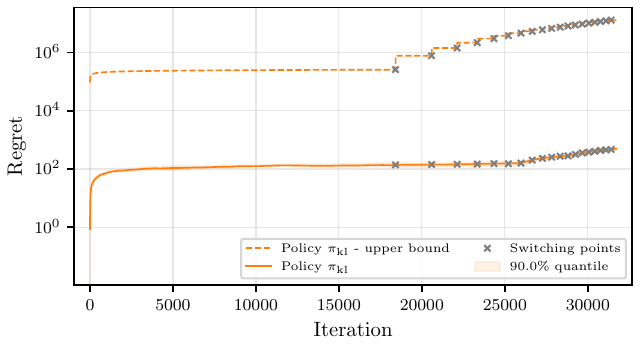}
    \caption{Comparison of the theoretical and simulated regret for $\klpolicy$.}
    \label{fig:kl_cb_regret}
\end{figure}

\subsection{Simulation Results for KL-Based Policy \klpolicy}
\begin{figure}[t]
    \centering
    \includegraphics[width=\imgscale\linewidth]{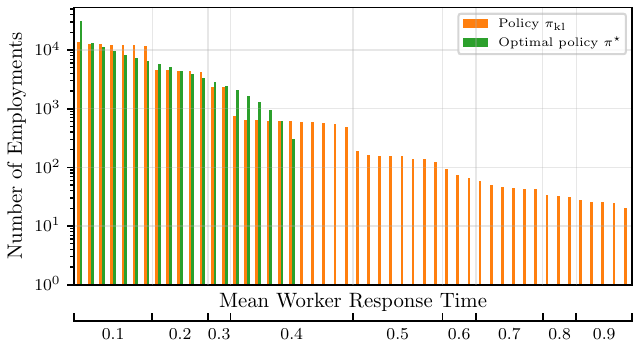}
    \caption{Number of worker employments for $\klpolicy$.}
    \label{fig:counters}
\end{figure}

The strategy proposed in \cref{sec:improved_policy} introduces additional computational overhead for the {\master} to run a numerical procedure for calculating the \acp{lcb} due to the missing analytical closed-form solution. Depending on the computational
resources of the main node, this overhead might outweigh
the benefits of using the KL-based policy. However, the convergence rate improves significantly compared to the strategy in \cref{sec:policy}. This is reflected by the regret in \cref{fig:kl_cb_regret}, where we find the regret bound again as a very pessimistic overestimate. The best workers in this case were determined with an accuracy of $99.0\%$, %
showing that the impacts of missclassfication can almost be neglected and the regret improves by a factor of approximately ten compared to the results in \cref{subsec:sim_res}. The improvement in regret follows directly from the reduced amount of suboptimal worker employments, which is depicted in \cref{fig:counters}.

\begin{remark}
One could try to find confidence bounds that further minimize the cumulative regret. However, in \cite{Zhong2021} it was shown that the two goals regret minimization and best arm identification become contradictory at some point. That is, one can either optimize an algorithm towards very confidently determining the best arms out of all available ones, or one could seek to optimize the cumulative regret to the maximum extend. Since in this case we are concerned with both objectives, the goal was to find a policy satisfying them simultaneously.
\end{remark}

\subsection{Comparison to Adaptive $k$-Sync}
\begin{figure}[t]
    \centering
    \includegraphics[width=\imgscale\linewidth]{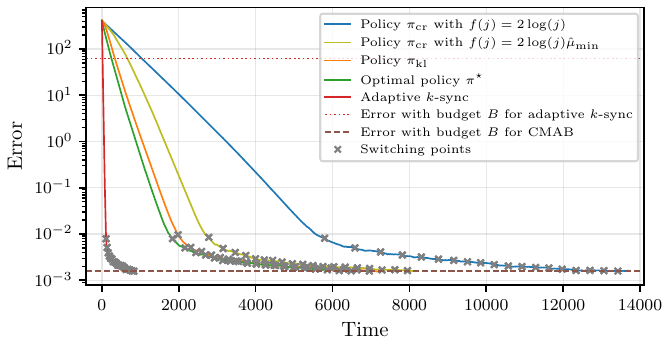}
    \caption{Comparison to adaptive $k$-sync \cite{Hanna2020} with limited budget $\totalbudget$.}
    \label{fig:comp_adaptive_k}
\end{figure}

In \cref{fig:comp_adaptive_k}, we analyze the convergence of the algorithms from \cref{sec:policy} (with two different $f(j)$) and \cref{sec:improved_policy}, and compare with the optimal policy $\policyopt$ and the adaptive $k$-sync scheme from \cite{Hanna2020}. %

\textbf{Speed.} As waiting for the $\roundidx$ fastest out of $\workercnt>\roundidx$ workers is on average faster than waiting for all out of $\roundidx$ workers, the adaptive $k$-sync strategy from \cite{Hanna2020} is significantly faster than our proposed scheme. Comparing $\policyopt$ to the performance of $\cmabpolicy$ with $\cbscale=2\log(j)$, learning the mean worker speeds slows down the convergence by a factor of almost three. This is because the chosen confidence radius mostly dominates the mean response time estimates of the workers, which leads to an emphasis on exploration, i.e., more confident estimates at the expense of sampling slow workers more often. While the confidence bound adapted by $\hat{\mu}_{\min}$ accounts for this drawback, $\klpolicy$ achieves the best results. However, this comes at the expense of more computational load to calculate the \acp{lcb}.

\textbf{Worker employment.} Considering the same cost, our proposed scheme is able to achieve significantly better results. In particular, with a total budget of $\totalbudget< \num{1.3e5}$ computations, the \ac{cmab}-based strategy reaches an error of $\approx \num{2e-3}$ while adaptive $k$-sync achieves an error of only $\approx \num{6e1}$.

\textbf{Communication.} The \ac{cmab} schemes had to transfer in total $\totalbudget < \num{1.3e5}$ models to the workers, while adaptive $k$-sync occupied the transmission link from the main node to the workers $\workercnt \cdot \timehorizonround[\budget] > \num{1.5e6}$ times. Thus, the downlink communication cost is reduced by more than a factor of ten, whereas the load on the uplink channel from the workers to the main node is $\totalbudget < \num{1.3e5}$ for both strategies. Consequently, the total amount of channel occupations has been reduced by more than $80\%$, i.e., $2\totalbudget$ instead of $\totalbudget + \workercnt \cdot \timehorizonround[\budget]$.

\section{Conclusion}\label{sec:conclusion}
In this paper, we have introduced a cost-efficient distributed machine learning scheme that assigns random tasks to fast workers and leverages all computations. The number of workers employed per iteration increases as the algorithm evolves. To speed up the convergence, we introduced the use of a \ac{cmab} model, for which we provided theoretical regret guarantees and simulation results. While our scheme is inferior to the adaptive $k$-sync strategy from \cite{Hanna2020} in terms of speed, it achieves much lower errors with the same computational efforts while reducing the communication load significantly. 

As further research directions, one could derive tighter regret bounds and improve the choice of the confidence bound. In addition, one can consider the setting in which the underlying distributions of the response times of the workers vary over time. Furthermore, relaxing the shared memory assumption and instead fixing the task distribution to the workers opens up an interesting trade-off between the average waiting time per iteration and the convergence rate for distributed machine learning with \acp{cmab}. On a high level, this holds because the main node should sample different subsets of the data at every iteration. Hence, the main node cannot always employ the fastest workers.

\bibliographystyle{IEEEtran}
\bibliography{IEEEabrv,refs}

\appendices

\crefalias{section}{appendix}

\section{Proof of \cref{prop:max_mean}} \label{sec:proof_prop_max_mean}
Let $F_\rvtmp$ be the cumulative distribution function of random variable $\rvtmp$, and let $\ps{\tmpsetsecond}$ be the power set of $\tmpsetsecond$. Consider exponentially distributed random variables indexed by a set $\tmpsetsecond$, i.e., $\rvtmp_\tmpvarthird \sim \exp\br{\lambda_\tmpvarthird}$, $\tmpvarthird \in \tmpsetsecond$, with different rates $\lambda_\tmpvarthird$ and cumulative distribution function $F_{\rvtmp_\tmpvarthird}(x) = 1 - e^{-\lambda_\tmpvarthird x}$. Then, we can derive the maximum order statistics, i.e., the expected value of the largest of their realizations, as
\begin{align} %
    \E{\max\limits_{\tmpvarthird \in \tmpsetsecond} \rvtmp_\tmpvarthird} &= \int\limits_0^\infty \left(1 - F_{\max\limits_{\tmpvarthird \in \tmpsetsecond} \rvtmp_\tmpvarthird}(x)\right) dx \nonumber \\
    &= \int\limits_0^\infty \br{1 - \prod\limits_{\tmpvarthird \in \tmpsetsecond} F_{\rvtmp_\tmpvarthird}(x)} dx \nonumber \\
    &= \int\limits_0^\infty \br{1 - \prod\limits_{\tmpvarthird \in \tmpsetsecond} (1 - e^{-\lambda_\tmpvarthird x})} dx \nonumber \\
    &= \int\limits_0^\infty \br{1 - \sum\limits_{\tmpsetthird \in \ps{\tmpsetsecond}} \br{-1}^{\vert \tmpsetthird \vert} e^{-\sum\limits_{\tmpvarforth \in \tmpsetthird} \lambda_\tmpvarforth x} dx} \nonumber \\
    &= \sum\limits_{\tmpsetthird \in \ps{\tmpsetsecond}\setminus\emptyset} \br{-1}^{\vert \tmpsetthird \vert - 1} \int\limits_0^\infty e^{-\sum\limits_{\tmpvarforth \in \tmpsetthird} \lambda_\tmpvarforth x} dx \nonumber,
\end{align}
Solving the integral concludes the proof.

\section{Proof of \cref{theorem:time_bound}} \label{sec:proof_time_bound}
To prove the probabilistic bound in \cref{theorem:time_bound}, we utilize the well-known \emph{Chebychev's inequality}. It provides a handle on the probability that a random variable $\rvtmp$ deviates from its mean by more than a given absolute value based on its variance. As given in \cite[p.~19]{Boucheron2013}, the relation denotes as
\begin{equation}
    \label{eq:chebyshev}
    \Prob{\card{\rvtmp - \E{\rvtmp} \geq \varepsilon}} \leq \frac{\Var{\rvtmp}}{\varepsilon^2}.
\end{equation}

The probability $1-\Pr(\iter, \roundidx)$ for a certain confidence parameter $\epsilon$ that the upper bound of the time spent in round $\roundidx$, i.e., $\mu_{\superarmbest} \!\!\br{\min\curly{\iter, \timehorizonround}-\timehorizonround[\roundidx-1]} \!\br{1+\epsilon}$, is smaller than the true run-time of the algorithm in iteration $\iter$, where $\iter > \timehorizonround[\roundidx-1]$, can be calculated by applying Chebychev's inequality given in \eqref{eq:chebyshev}:
\ifdouble\begin{align*}
    1\!-\!\Pr(\iter, \roundidx) = \, 
    &\begin{aligned}[t]
        \Pr\Bigg(\sum\limits_{\tmpvar = \timehorizonround[\roundidx-1]+1}^{\min\curly{\iter, \timehorizonround}}\!\!\!\! &\superarmbestresp[\tmpvar] \!\! - \! \mu_{\superarmbest} \! \br{\min\curly{\iter, \timehorizonround}-\timehorizonround[\roundidx-1]} \\&\geq \epsilon \cdot \mu_{\superarmbest} \br{\min\curly{\iter, \timehorizonround}-\timehorizonround[\roundidx-1]} \vphantom{\sum\limits_{\tmpvar = \timehorizonround[\roundidx-1]+1}^{\min\curly{\iter, \timehorizonround}}}\Bigg)
    \end{aligned}\\ 
    &\leq \frac{\Var{\sum\limits_{\tmpvar = \timehorizonround[\roundidx-1]+1}^{\min\curly{\iter, \timehorizonround}} \superarmbestresp[\tmpvar]}}{\mu_{\superarmbest}^2 \br{\min\curly{\iter, \timehorizonround}-\timehorizonround[\roundidx-1]}^2 \epsilon^2} \\
    &\leq \frac{\sigma_{\superarmbest}^2}{\mu_{\superarmbest}^2 \br{\min\curly{\iter, \timehorizonround}-\timehorizonround[\roundidx-1]} \epsilon^2}
\end{align*}\else
\begin{align*}
    1\!-\!\Pr(\iter, \roundidx) = \, 
    &\begin{aligned}[t]
        \Pr\Bigg(\sum\limits_{\tmpvar = \timehorizonround[\roundidx-1]+1}^{\min\curly{\iter, \timehorizonround}}\!\!\!\! &\superarmbestresp[\tmpvar] \!\! - \! \mu_{\superarmbest} \! \br{\min\curly{\iter, \timehorizonround}-\timehorizonround[\roundidx-1]} \geq \epsilon \cdot \mu_{\superarmbest} \br{\min\curly{\iter, \timehorizonround}-\timehorizonround[\roundidx-1]} \vphantom{\sum\limits_{\tmpvar = \timehorizonround[\roundidx-1]+1}^{\min\curly{\iter, \timehorizonround}}}\Bigg)
    \end{aligned}\\ 
    &\leq \frac{\Var{\sum\limits_{\tmpvar = \timehorizonround[\roundidx-1]+1}^{\min\curly{\iter, \timehorizonround}} \superarmbestresp[\tmpvar]}}{\mu_{\superarmbest}^2 \br{\min\curly{\iter, \timehorizonround}-\timehorizonround[\roundidx-1]}^2 \epsilon^2} \\
    &\leq \frac{\sigma_{\superarmbest}^2}{\mu_{\superarmbest}^2 \br{\min\curly{\iter, \timehorizonround}-\timehorizonround[\roundidx-1]} \epsilon^2}
\end{align*}\fi

Then, the probability that $\timevar_\iter$ is underestimated in any of the rounds up to iteration $\iter$ eventually can be given as\footnote{Please note that $\Pr(j)$ and $\Pr(j,r)$ correspond to the desired events, i.e., the probabilities that the true run-time of the algorithm is less than or equal to the upper bound.}
\begin{align*}
    1\! -\! \Pr(\iter) \! &= \! 1 - \!\!\!\!\!\!\!\!\!\!\!\prod\limits_{r \in \sq{1, \budget}: \iter > \timehorizonround[\roundidx-1]} \!\! \br{1- (1-\Pr(\iter, \roundidx))} \\
    &\leq 1 - \!\!\!\!\!\!\!\!\!\!\!\prod\limits_{r \in \sq{1, \budget}: \iter > \timehorizonround[\roundidx-1]} \!\!\!\! \br{\!1 - \frac{\sigma_{\superarmbest}^2}{\mu_{\superarmbest}^2 \br{\min\curly{\iter, \timehorizonround}-\timehorizonround[\roundidx-1]} \epsilon^2}\!}\!.
\end{align*}
Taking the complementary event concludes the proof.

\section{Well-Known Tail Bounds} \label{appendix:bounds}

In \cref{sec:proof}, we utilize well-known properties of tail distributions of sub-gamma and sub-Gaussian random variables, which we provide in the following for completeness.

\emph{Sub-gamma tail bound:}
For $\rvtmp \sim \subgamma{\alpha}{\beta}$ with variance $\sigma^2=\frac{\alpha}{\beta^2}$ and scale $c=\frac{1}{\beta}$, we have that \cite[p.~29]{Boucheron2013}
$
  \label{eq:subgamma_tail_bound_eps}
  \Prob{\rvtmp > \sqrt{2 \sigma^2 \varepsilon} + c \varepsilon} \leq \exp\br{-\varepsilon}.
$

\emph{Sub-Gaussian tail bound:}
Resulting from the Cramér-Chernoff method, we obtain for any $\rvtmp \sim \subgaussian{\sigma^2}$ \cite[p.~77]{Lattimore2020} that
$
\label{eq:subgaussian_tail_bound_eps}
P\br{\rvtmp \leq -\varepsilon} \leq \exp\br{-\frac{\varepsilon^2}{2\sigma^2}}.
$

\section{Proof of \cref{theorem:regret_bound}} \label{sec:proof}

While we will benefit from the proof strategies in \cite{Auer2002} and \cite{Gai2012}, our analysis differs in that we consider an unbounded distribution of the rewards, which requires us to investigate the properties of sub-gamma distributions and to use different confidence bounds. Also, compared to \cite{Auer2002}, we deal with \acp{lcb} instead of \acp{ucb} since we want to minimize the response time of a superarm, i.e., the time spent per iteration. This problem setting was briefly discussed in \cite{Gai2012}. While the authors bound the probability of overestimating an entire suboptimal superarm in \cite{Gai2012}, we bound the probability of individual suboptimal arm choices. This is justified by independent outcomes across arms and by the combined outcome of a superarm being a monotonically non-decreasing function of the individual arms' rewards, that is, the workers' mean response times. %
A superarm $\superarm$ is considered suboptimal if $\mu_{\superarm} > \mu_{\superarmbest}$ and a single arm $\superarmelement$ is suboptimal if $\mu_{\superarmelement} > \mu_{\superarmbestelement}$.
In addition to the counter $\pullcounter$, for every arm $\armidx \in [\workercnt]$, we introduce the counter $\failcountermax \leq \pullcounter$. %
The integer $\cardmax$ refers to the maximum cardinality of all possible superarm choices, i.e., $\failcountermax$ is valid for all rounds $\roundidx \leq \cardmax$. If a suboptimal superarm $\superarm$ is chosen\footnote{Although there exists an optimal superarm, it is not necessarily unique, i.e., there might exist several superarms $\superarm$ with $\mu_{\superarm} = \mu_{\superarmbest}$.}, $\failcountermax$ is incremented only for the arm $\armidx\in \superarm$ that has been pulled the least until this point in time, i.e., $\armidx = \superarmvar^\roundidx_{\element_{\text{min}}} (\iter)$, where $\element_{\text{min}} = \argmin_{\element \in \superarm} \pullcountervar{\element}$. Hence, $\sum_{\armidx=1}^\armcnt \failcountermax$ equals the number of suboptimal superarm pulls. Let $\superarmspacecardlim$ be the set of all superarms with a maximum cardinality of $\cardmax$, i.e., $\superarmspacecardlim\define \bigcup_{\roundidx \leq \cardmax} \superarmspace$, and $\pullcountervar{\superarmvar}$ the number of times superarm $\superarmvar$ has been pulled until iteration $\iter$. We have %
\begin{equation} \label{eq:counter_interdep}
    \sum\limits_{\tmpset \in \superarmspacecardlim:\mu_{\tmpset} > \mu_{\superarmvar^{\card{\tmpset},\star}}} \E{\pullcountervar{\tmpset}} = \sum\limits_{\armidx=1}^\armcnt \E{\failcountermax}.
\end{equation}

Applying \cite[Lemma 4.5]{Lattimore2020} to express the regret in terms of the suboptimality gaps in iteration $\iter$ of round $\roundidx$, i.e., $\timehorizonround[\roundidx-1] < \iter \leq \timehorizonround[\roundidx]$ and $\cardmax = \roundidx$, we use \eqref{eq:counter_interdep} and obtain 
\begin{align}
    R_{\timeidx}^{\cmabpolicy} %
    &= \sum\limits_{\tmpset \in \superarmspacecardlim:\mu_{\tmpset} > \mu_{\superarmvar^{\card{\tmpset},\star}}} \Delta_\tmpset \, \E{\pullcountervar{\tmpset}} \nonumber \\
    &\leq \max\limits_{\roundidx \in \sq{1, \budget}: \timeidx>\timehorizon_{\roundidx-1}} \Delta_{\superarmvar^{\roundidx}, \max} \cdot \sum\limits_{\armidx=1}^\armcnt \E{\failcountermax}. \label{eq:3}
\end{align} %
To conclude the proof of~\cref{theorem:regret_bound} we need the following intermediate results. %

\begin{lemma} \label{claim:bound_suboptimal_choice}
Let $\card{\superarm[\tmpvar]} \leq \cardmax$ hold $\forall \tmpvar \in \sq{\iter}$. For any $h\geq 0$,
we can bound the expectation of $\failcountermax$, $\armidx \in \sq{\workercnt}$, as
\ifdouble\begin{align*}
\E{\failcountermax} \leq 
    & \,h + 
    \sum\limits_{\tmpvar = 1}^{\timeidx} 
    \timeidx^{2} \cdot 
    \sum\limits_{\element=1}^{\card{\superarm[\tmpvar]}}
    \Pr \left(\mu_{\superarmelement[\tmpvar]} > \mu_{\superarmbestelement}, \right. \\
    & \left. \cbshort[\tmpvar]{\superarmelement[\tmpvar]} \leq
    \cbshort[\tmpvar]{\superarmbestelement} \right).
\end{align*}\else
\begin{align*}
\E{\failcountermax} \leq 
    & \,h + 
    \sum\limits_{\tmpvar = 1}^{\timeidx} 
    \timeidx^{2} \cdot 
    \sum\limits_{\element=1}^{\card{\superarm[\tmpvar]}}
    \Pr \left(\mu_{\superarmelement[\tmpvar]} > \mu_{\superarmbestelement}, \right. \left. \cbshort[\tmpvar]{\superarmelement[\tmpvar]} \leq
    \cbshort[\tmpvar]{\superarmbestelement} \right).
\end{align*}\fi
\end{lemma}

By construction, for a superarm $\superarm$ with $\card{\superarm} \leq \cardmax$ it holds for all $\element \in [1, \card{\superarm}]$ and $\armidx \in \superarm$ that $\pullcountervar{\superarmelement} \geq \failcountermax$. By applying \cref{claim:bound_suboptimal_choice} with $h \geq 0$, we thus have $\pullcountervar{\superarmelement} \geq \failcountermax \geq \counteroffset$. To bound the probability of choosing a suboptimal arm over which we sum in \cref{claim:bound_suboptimal_choice}, we use \cref{lemma:wrong_estimate}.

\begin{lemma} \label{lemma:wrong_estimate}
The probability of the $\element$-th fastest arm of $\superarm$ being suboptimal given that $\pullcountervar{\superarmelement} \geq \lceil\roundoffset\rceil$, is bounded from above as
\ifdouble\begin{align}
    &\Prob{\mu_{\superarmelement} > \mu_{\superarmbestelement}, \cbshort{\superarmelement} \leq
    \cbshort{\superarmbestelement}} \nonumber \\
    &\leq \timeidx^{-4 \min\curly{\lambda_{\min}, \lambda_{\min}^2}} + \timeidx^{-4 \lambda_{\min}}.
\end{align}\else
\begin{align}
    &\Prob{\mu_{\superarmelement} > \mu_{\superarmbestelement}, \cbshort{\superarmelement} \leq
    \cbshort{\superarmbestelement}} \nonumber 
    \leq \timeidx^{-4 \min\curly{\lambda_{\min}, \lambda_{\min}^2}} + \timeidx^{-4 \lambda_{\min}}.
\end{align}\fi
\end{lemma}

Having the results of~\cref{claim:bound_suboptimal_choice} and~\cref{lemma:wrong_estimate}, and choosing $h =\lceil\roundoffset\rceil$, for $\lambda_{\min}\geq1$ we have
\ifdouble\begin{align}
    \E{\failcountermax} &\leq
    h + 
    \sum\limits_{\tmpvar = 1}^{\timeidx} 
    \timeidx^{2} \cdot 
    \sum\limits_{\element=1}^{\card{\superarm[\tmpvar]}}
    \Pr \left(\mu_{\superarmelement[\tmpvar]} > \mu_{\superarmbestelement}, \right. \nonumber \\
    &\left. \cbshort[\tmpvar]{\superarmelement[\tmpvar]} \leq
    \cbshort[\tmpvar]{\superarmbestelement} \right) \nonumber \\
    &\leq
    \ceil{\roundoffset} + 
    \sum\limits_{\tmpvar = 1}^{\timeidx} 
    \timeidx^{2} \cdot
    \sum\limits_{\element=1}^\cardmax
    2\timeidx^{-4 \lambda_{\min}} \nonumber \\
    &\leq
    \roundoffsetbounded + 1 +
    \cardmax \cdot \frac{\pi}{3},\label{eq:lemma4}
\end{align}\else
\begin{align}
    \E{\failcountermax} &\leq
    h + 
    \sum\limits_{\tmpvar = 1}^{\timeidx} 
    \timeidx^{2} \cdot 
    \sum\limits_{\element=1}^{\card{\superarm[\tmpvar]}}
    \Pr \left(\mu_{\superarmelement[\tmpvar]} > \mu_{\superarmbestelement}, \right. \nonumber 
    \left. \cbshort[\tmpvar]{\superarmelement[\tmpvar]} \leq
    \cbshort[\tmpvar]{\superarmbestelement} \right) \nonumber \\
    &\leq
    \ceil{\roundoffset} + 
    \sum\limits_{\tmpvar = 1}^{\timeidx} 
    \timeidx^{2} \cdot
    \sum\limits_{\element=1}^\cardmax
    2\timeidx^{-4 \lambda_{\min}} \nonumber \\
    &\leq
    \roundoffsetbounded + 1 +
    \cardmax \cdot \frac{\pi}{3},\label{eq:lemma4}
\end{align}\fi
where the last step relates to the Basel problem of a $p$-series\footnote{We need $\lambda_{\min}\geq 1$ so that the $p$-series converges to a small value.}. Plugging the bound in~\eqref{eq:lemma4} into~\eqref{eq:3} concludes the proof. 

The proof of~\cref{claim:bound_suboptimal_choice} uses standard techniques from the literature on \acp{mab} and is given in the following for completeness. 
The proof of~\cref{lemma:wrong_estimate} is given afterwards.
\begin{proof}[Proof of~\cref{claim:bound_suboptimal_choice}]
At first, we bound the counter $\failcountermax$ from below by introducing an arbitrary parameter $\counteroffset$ that eventually serves to limit the probability of choosing a suboptimal arm. Similar to \cite{Gai2012}, we have
\ifdouble\begin{alignat}{2}
    &C_{i,e} (j) \leq
    \begin{aligned}[t]
        &\counteroffset + 
        \sum\limits_{\tmpvar = 1}^{\timeidx} 
        \mathds{1} \left\{ \exists\; 1 \leq \element \leq \card{\superarm[\tmpvar]}: \right.\left. \mu_{\superarmelement[\tmpvar]} > \mu_{\superarmbestelement}, \right. \nonumber \\
        &\left. \cbshort[\tmpvar]{\superarmelement[\tmpvar]} \right.\left.\leq \cbshort[\tmpvar]{\superarmbestelement} \right\}
    \end{aligned}
     \nonumber \\
    &\,\,\,\,\leq 
    \begin{aligned}[t]
        &\counteroffset + 
        \sum\limits_{\tmpvar = 1}^{\timeidx} 
        \sum\limits_{\element=1}^{\card{\superarm[\tmpvar]}} \mathds{1} \{\mu_{\superarmelement[\tmpvar]} > \mu_{\superarmbestelement}, \nonumber \\ &\cbshort[\tmpvar]{\superarmelement[\tmpvar]} \leq
        \cbshort[\tmpvar]{\superarmbestelement} \} 
        \end{aligned} 
    \nonumber \\
    &\,\,\,\,\leq \begin{aligned}[t]
    &\counteroffset + 
    \sum\limits_{\tmpvar = 1}^{\timeidx} 
    \sum\limits_{\element=1}^{\card{\superarm[\tmpvar]}}
    \mathds{1} \{\mu_{\superarmelement[\tmpvar]} > \mu_{\superarmbestelement}, 
    \nonumber \\ &\min\limits_{\counteroffset \leq \pullcountervar[\tmpvar]{\superarmelement[\tmpvar]} \leq \timeidx} \cbshort[\tmpvar]{\superarmelement[\tmpvar]} \leq
    \max\limits_{1 \leq \pullcountervar[\tmpvar]{\superarmbestelement} \leq \timeidx} \cbshort[\tmpvar]{\superarmbestelement} \} \end{aligned} \nonumber \\
    &\,\,\,\,\leq \begin{aligned}[t]
    &\counteroffset + 
    \sum\limits_{\tmpvar = 1}^{\timeidx} 
    \sum\limits_{\pullcountervar[\tmpvar]{\superarmelement[\tmpvar]} = \counteroffset}^\timeidx
    \sum\limits_{\pullcountervar[\tmpvar]{\superarmbestelement} = 1}^\timeidx \sum\limits_{\element=1}^{\card{\superarm[\tmpvar]}} \mathds{1} \{\mu_{\superarmelement[\tmpvar]} > \mu_{\superarmbestelement}, \nonumber \\
    &\cbshort[\tmpvar]{\superarmelement[\tmpvar]} \leq
    \cbshort[\tmpvar]{\superarmbestelement}\} \end{aligned} \nonumber \\
    &\,\,\,\,\leq \begin{aligned}[t]
    &\counteroffset + 
    \sum\limits_{\tmpvar = 1}^{\timeidx} 
    \iter^2 \sum\limits_{\element=1}^{\card{\superarm[\tmpvar]}} \mathds{1} \{\mu_{\superarmelement[\tmpvar]} > \mu_{\superarmbestelement},  \\ &\cbshort[\tmpvar]{\superarmelement[\tmpvar]} \leq
    \cbshort[\tmpvar]{\superarmbestelement}\}, \end{aligned} \nonumber %
\end{alignat}\else
\begin{alignat}{2}
    C_{i,e} (j) &\leq
    \begin{aligned}[t]
        &\counteroffset + 
        \sum\limits_{\tmpvar = 1}^{\timeidx} 
        \mathds{1} \left\{ \exists\; 1 \leq \element \leq \card{\superarm[\tmpvar]}: \right.\left. \mu_{\superarmelement[\tmpvar]} > \mu_{\superarmbestelement}, \right. \nonumber \left. \cbshort[\tmpvar]{\superarmelement[\tmpvar]} \right.\left.\leq \cbshort[\tmpvar]{\superarmbestelement} \right\}
    \end{aligned}
     \nonumber \\
    &\leq 
    \begin{aligned}[t]
        &\counteroffset + 
        \sum\limits_{\tmpvar = 1}^{\timeidx} 
        \sum\limits_{\element=1}^{\card{\superarm[\tmpvar]}} \mathds{1} \{\mu_{\superarmelement[\tmpvar]} > \mu_{\superarmbestelement}, \nonumber \cbshort[\tmpvar]{\superarmelement[\tmpvar]} \leq
        \cbshort[\tmpvar]{\superarmbestelement} \} 
        \end{aligned} 
    \nonumber \\
    &\leq \begin{aligned}[t]
    &\counteroffset + 
    \sum\limits_{\tmpvar = 1}^{\timeidx} 
    \sum\limits_{\element=1}^{\card{\superarm[\tmpvar]}}
    \mathds{1} \{\mu_{\superarmelement[\tmpvar]} > \mu_{\superarmbestelement}, 
    \nonumber \min\limits_{\counteroffset \leq \pullcountervar[\tmpvar]{\superarmelement[\tmpvar]} \leq \timeidx} \cbshort[\tmpvar]{\superarmelement[\tmpvar]} \leq
    \max\limits_{1 \leq \pullcountervar[\tmpvar]{\superarmbestelement} \leq \timeidx} \cbshort[\tmpvar]{\superarmbestelement} \} \end{aligned} \nonumber \\
    &\leq \begin{aligned}[t]
    &\counteroffset + 
    \sum\limits_{\tmpvar = 1}^{\timeidx} 
    \sum\limits_{\pullcountervar[\tmpvar]{\superarmelement[\tmpvar]} = \counteroffset}^\timeidx
    \sum\limits_{\pullcountervar[\tmpvar]{\superarmbestelement} = 1}^\timeidx \sum\limits_{\element=1}^{\card{\superarm[\tmpvar]}} \mathds{1} \{\mu_{\superarmelement[\tmpvar]} > \mu_{\superarmbestelement}, \nonumber \cbshort[\tmpvar]{\superarmelement[\tmpvar]} \leq
    \cbshort[\tmpvar]{\superarmbestelement}\} \end{aligned} \nonumber \\
    &\leq \begin{aligned}[t]
    &\counteroffset + 
    \sum\limits_{\tmpvar = 1}^{\timeidx} 
    \iter^2 \sum\limits_{\element=1}^{\card{\superarm[\tmpvar]}} \mathds{1} \{\mu_{\superarmelement[\tmpvar]} > \mu_{\superarmbestelement}, \cbshort[\tmpvar]{\superarmelement[\tmpvar]} \leq
    \cbshort[\tmpvar]{\superarmbestelement}\}, \end{aligned} \nonumber %
\end{alignat}\fi
where the first line describes the event of choosing a suboptimal superarm in terms of the events that the \ac{lcb} of any suboptimal $\element$-fastest arm in $\superarm$ is less than or equal to the \ac{lcb} of the $\element$-fastest arm in $\superarmbest$. %
To conclude the proof, we take the expectation on both sides. %
\end{proof}

\begin{proof}[Proof of \cref{lemma:wrong_estimate}]
As given in \cite{Auer2002}, to overestimate the $\element$-th fastest arm of $\superarm$, i.e., for $\cbshort{\superarmelement} \leq \cbshort{\superarmbestelement}$ to hold,
at least one of the following events must be satisfied:
\begin{align}
  \mu_{\superarmbestelement} & > \mu_{\superarmelement} - 2\cbradshort{\superarmelement}{\iter} \label{eq:req3_opt_diff},\\
    \muhat{\superarmbestelement} &\geq \mu_{\superarmbestelement} + \cbradshort{\superarmbestelement}{\iter} \label{eq:req1_right_tail}, \\
    \muhat{\superarmelement} &\leq \mu_{\superarmelement} - \cbradshort{\superarmelement}{\iter} \label{eq:req2_left_tail}.
\end{align}
    
Let in the following $\cbscale \define 2\log(\timeidx)$. We first show that the requirement\footnote{The scaling factor $\gamma = 24$ is chosen as an approximation of the exact solution of $2\br{\sqrt{\frac{4}{\gamma}} + \frac{2}{\gamma}} = 1$, which is $4\br{3 + 2\sqrt{2}} \approx 23.31 \leq 24$.} $\pullcountervar{\superarmelement} \geq \lceil\frac{24 f(\timeidx)}{\min\{\delta_{\min}^2, \delta_{\min}\}}\rceil \define \ell$ guarantees that $\mu_{\superarmbestelement} - \mu_{\superarmelement} + 2\cbradshort{\superarmelement}{\iter} \leq 0$, for all $\superarm \in \superarmspace$ with $\roundidx\in [1, \budget]$, $\element\in [1,\roundidx]$ and $\mu_{\superarmelement} > \mu_{\superarmbestelement}$, thus making the event~\eqref{eq:req3_opt_diff} a zero-probability event. We have
\begin{align}
    \mu_{\superarmbestelement}&  - \mu_{\superarmelement} + 2\br{\cbradcomb{\pullcountervar{\superarmelement}}} \nonumber \\
    & \leq \mu_{\superarmbestelement} - \mu_{\superarmelement} + 2\br{\cbradcomb{\counteroffset}} \nonumber \\
    & \leq -\delta_{\superarmelement} + \delta_{\min} \leq 0. \nonumber
\end{align}
\begin{lemma} \label{lemma:deviation_dist}
Given i.i.d. random variables $\armresp \sim \exp(\lambda_\armidx)$, $\iter = 1, \dots, \timehorizon$, the deviation of the empirical mean from the true mean $\muihat - \mui \define \frac{1}{\timehorizon} \sum_{\iter=1}^\timehorizon \left(\armresp - \mathbb{E}[\armresp]\right)$ follows a sub-gamma distribution $\subgamma{\timehorizon}{\timehorizon \lambda_\armidx}$ on the right tail and a sub-Gaussian distribution $\subgaussian{\frac{1}{T\lambda_\armidx^2}}$ on the left tail.
\end{lemma}
We apply \cref{lemma:deviation_dist} (which is proven below) to bound the probability of the event \eqref{eq:req1_right_tail}. However, two cases have to be distinguished. For $\lambda_{\min} \geq 1$, we have
\begin{align}
    &\Prob{\muhat{\superarmbestelement} \geq \mu_{\superarmbestelement} + \cbradshort{\superarmbestelement}{\iter}} \nonumber \\
    &= \Prob{\muhat{\superarmbestelement} \geq \mu_{\superarmbestelement} +
    \sqrt{\frac{4 \cbscale}{\pullcountervar{\superarmbestelement}}} + \frac{2 \cbscale}{\pullcountervar{\superarmbestelement}}} \nonumber \\
    &\leq \Prob{\muhat{\superarmbestelement} \geq \mu_{\superarmbestelement} +
    \sqrt{\frac{4 \cbscale}{\pullcountervar{\superarmbestelement} \lambda_{\min}}} + \frac{2 \cbscale}{\pullcountervar{\superarmbestelement}}} \nonumber \\
    &\leq \Prob{\muhat{\superarmbestelement} \geq \mu_{\superarmbestelement} \! +\!
    \sqrt{\frac{4 \cbscale \lambda_{\min}}{\pullcountervar{\superarmbestelement} \! \lambda_{\min}^2}} \! + \! \frac{2 \cbscale \lambda_{\min}}{\pullcountervar{\superarmbestelement} \! \lambda_{\min}}} \nonumber \\
    &= \Prob{\muhat{\superarmbestelement} \geq \mu_{\superarmbestelement} \!+ \!
    \sqrt{\frac{8 \log(\timeidx) \lambda_{\min}^2}{\pullcountervar{\superarmbestelement} \lambda_{\min}^2}} + \frac{4 \log(\timeidx) \lambda_{\min}^2}{\pullcountervar{\superarmbestelement} \lambda_{\min}}} \nonumber \\
    &\leq \exp{-4\log(\timeidx)\lambda_{\min}} \leq \timeidx^{-4 \lambda_{\min}}, \nonumber
\end{align}
where in the penultimate step we used the sub-gamma tail bound given in \cref{appendix:bounds} with $\varepsilon = 2\log(\timeidx)\lambda_{\min}$.

In contrast, if $\lambda_{\min} < 1$
\begin{align*}
    &\Prob{\muhat{\superarmbestelement} \geq \mu_{\superarmbestelement} + \cbradshort{\superarmbestelement}{\iter}} \nonumber \\
    &= \Prob{\muhat{\superarmbestelement} \geq \mu_{\superarmbestelement} + \cbradcomb{\pullcountervar{\superarmbestelement}}} \nonumber \\
    &\leq \Prob{\muhat{\superarmbestelement} \geq \mu_{\superarmbestelement} +
    \sqrt{\frac{4 \cbscale}{\pullcountervar{\superarmbestelement}}} + \frac{2 \cbscale \lambda_{\min}}{\pullcountervar{\superarmbestelement}}} \nonumber \\
    &= \Prob{\muhat{\superarmbestelement} \geq \mu_{\superarmbestelement} +
    \sqrt{\frac{4 \cbscale \lambda_{\min}^2}{\pullcountervar{\superarmbestelement} \lambda_{\min}^2}} + \frac{2 \cbscale \lambda_{\min}^2}{\pullcountervar{\superarmbestelement} \lambda_{\min}}} \nonumber \\
    &= \Prob{\muhat{\superarmbestelement} \geq \mu_{\superarmbestelement} +
    \sqrt{\frac{8 \log(\timeidx) \lambda_{\min}^2}{\pullcountervar{\superarmbestelement} \lambda_{\min}^2}} + \frac{4 \log(\timeidx) \lambda_{\min}^2}{\pullcountervar{\superarmbestelement} \lambda_{\min}}} \nonumber \\
    &\leq \exp{-4 \log(\timeidx)\lambda_{\min}^2} \nonumber \\
    &\leq \timeidx^{-4 \lambda_{\min}^2},
\end{align*}
where in the penultimate line we used the sub-gamma tail bound in \cref{appendix:bounds} with $\varepsilon = 2\log(\timeidx)\lambda_{\min}^2$.

For the event in \eqref{eq:req2_left_tail}, we have %
\begin{align}
    &\Prob{\muhat{\superarmelement} \leq \mu_{\superarmelement} - \cbradshort{\superarmelement}{\iter}} \nonumber \\
    &= \Prob{\muhat{\superarmelement} \leq \mu_{\superarmelement} - \sqrt{\frac{4 \cbscale}{\pullcountervar{\superarmelement}}} - \frac{2 \cbscale}{\pullcountervar{\superarmelement}}} \nonumber \\
    &\leq \Prob{\muhat{\superarmelement} \leq \mu_{\superarmelement} - \sqrt{\frac{4 \cbscale}{\pullcountervar{\superarmelement}}}} \nonumber \\
    &\leq \exp{-4\lambda_{\superarmelement}^2 \log(\timeidx)} \nonumber \\
    &\leq \timeidx^{-4\lambda_{\superarmelement}^2} \leq \timeidx^{-4\lambda_{\min}^2} \leq \timeidx^{-4\lambda_{\min}}, \nonumber
\end{align}
where we used the sub-Gaussian tail bound given in \cref{appendix:bounds}.
\end{proof}

\begin{proof}[Proof of \cref{lemma:deviation_dist}]
To conduct the proof, we express the independently and identically distributed random variables $\armresp \sim \exp(\lambda_\armidx)$, $\iter \in [1, \timehorizon]$ in terms of its moment generating functions $M_{\rvtmp_i}(\mgfarg) = \E{e^{\mgfarg \rvtmp_i}} = \frac{\lambda_\armidx}{\lambda_\armidx-\mgfarg}$, which is defined for $\mgfarg<\lambda$. Summing the identically distributed realizations is equivalent to multiplying its moment generating functions, i.e., $\sum_{\timeidx=1}^{\timehorizon} \armresp \leftrightarrow \prod_{\timeidx=1}^{\timehorizon} M_{\armresp}(\mgfarg) = \br{\frac{\lambda_\armidx}{\lambda_\armidx -\mgfarg}}^\timehorizon$, which describes a random variable related to a gamma distribution with shape $\timehorizon$ and rate $\lambda_\armidx$. The empirical mean $\muihat = \frac{1}{\timehorizon} \sum_{\timeidx=1}^\timehorizon \armresp$ is thus distributed according to the scaled gamma distribution $\Gamma(\timehorizon, \lambda_\armidx \timehorizon)$ with mean $\mu_\armidx = \frac{1}{\lambda_\armidx}$ and variance $\sigma^2=\frac{1}{T \lambda_\armidx^2}$. Thus, $\muihat - \mui$ is a centered gamma distributed random variable, which according to \cite[pp.~27]{Boucheron2013} has the properties stated in \cref{lemma:deviation_dist}.
\end{proof}

\section{Proof of \cref{theorem:kl_regret}} \label{sec:proof_kl}

In the following, we prove the regret bounds given in \cref{theorem:kl_regret} based on the analysis in \cite{Garivier2013}. In particular, we adapt the regret proof in \cite{Garivier2013} for single arm choices to prove a combinatorial regret bound. To start with, we use the same counter definitions as in \cref{sec:proof} and follow the same steps, i.e., we apply \cite[Lemma 4.5]{Lattimore2020} and use the interdependencies between the counters, to finally result in \eqref{eq_line:kl_regret_prefinal}. Let  $\cardmax = \max\limits_{\roundidx \in \sq{1, \budget}: \timeidx>\timehorizon_{\roundidx-1}} \roundidx$, then
\begin{align}
    \regret 
    &\leq \max\limits_{\roundidx \in \sq{1, \budget}: \timeidx>\timehorizon_{\roundidx-1}} \Delta_{\superarmvar^{\roundidx}, \max} \cdot \sum\limits_{\armidx=1}^\armcnt \E{\failcountermax}. \label{eq_line:kl_regret_prefinal}
\end{align}
To conclude, we need to bound the expectation of $\failcountermax$. A handle on this is given in the following \cref{lemma:kl_failcounter_bound}.
\begin{lemma} \label{lemma:kl_failcounter_bound} The expectation of the suboptimal arm counter $\failcountermax$ for superarms $\superarm[\tmpvar]$, $\tmpvar \in \sq{\iter}$, with maximum cardinality $\cardmax$, i.e., $\forall \tmpvar \in \sq{1, \iter}$, $\card{\superarm[\roundidx]} \leq \cardmax$, \acp{lcb} according to \eqref{eq:kl_cb}, and $f(\timeidx) = \log(\timeidx) + 3\log\br{\log\br{\timeidx}}$, $\iter>3$, can be bounded as
\begin{align*}
    \E{\failcountermax} &\leq
    \cardmax \br{7 \log(\log(\timeidx)) + \frac{1 + \epsilon}{\klmin} f(\timeidx) + Q_\iter(\epsilon)},
\end{align*}
where $Q_\iter(\epsilon) = \frac{\exp\br{-\klepsmin \br{\frac{1 + \epsilon}{\klmax} f(\timeidx) - 1}}}{1 - \exp(-\klepsmin)}$.
\end{lemma}

Plugging the result of \cref{lemma:kl_failcounter_bound} into \eqref{eq_line:kl_regret_prefinal} concludes the proof of \cref{theorem:kl_regret}. %
It remains to prove \cref{lemma:kl_failcounter_bound}.

\begin{proof}[Proof of \cref{lemma:kl_failcounter_bound}]
To prove \cref{lemma:kl_failcounter_bound}, we need a result from \cite{Garivier2013}, which we state in our notation and for our problem in \cref{lemma:kl_suboptimal_total_prob_bound}. We briefly explain why this result holds in our setting.
\begin{lemma}[\!\!\cite{Garivier2013}] \label{lemma:kl_suboptimal_total_prob_bound}
The probability of overestimating the $\element$-th fastest element of $\superarm$ in any of the iterations up to iteration $\iter$ given that the element is suboptimal, %
a decision process $\klpolicy$ based on the \ac{lcb} in \eqref{eq:kl_cb} and $K_\timeidx = \Big\lfloor \frac{1 + \epsilon}{\klminus{\mu_{\superarmelement}}{\mu_{\superarmbestelement}}} f(\timeidx) \Big\rfloor$ can be upper bounded as 
\ifdouble\begin{align*}
    \sum\limits_{\tmpvar = 1}^{\iter} &\Prob{\mu_{\superarmelement[\tmpvar]} > \mu_{\superarmbestelement}, \cbshort[\tmpvar]{\superarmelement[\tmpvar]} \leq \cbshort[\tmpvar]{\superarmbestelement}} \\
    &\leq 7 \log(\log(\timeidx)) + \frac{1 + \epsilon}{\klminus{\mu_{\superarmelement}}{\mu_{\superarmbestelement}}} f(\timeidx) \\ &+\frac{\exp\br{-\kl{\phi(\epsilon, \mu_{\superarmbestelement}, \mu_{\tmpset_\element})}{\mu_{\superarmelement}} K_\timeidx}}{1 - \exp(-\kl{\phi(\epsilon, \mu_{\superarmbestelement}, \mu_{\tmpset_\element})}{\mu_{\superarmelement}}},
\end{align*}\else
\begin{align*}
    \sum\limits_{\tmpvar = 1}^{\iter} &\Prob{\mu_{\superarmelement[\tmpvar]} > \mu_{\superarmbestelement}, \cbshort[\tmpvar]{\superarmelement[\tmpvar]} \leq \cbshort[\tmpvar]{\superarmbestelement}} \\
    &\leq 7 \log(\log(\timeidx)) + \frac{1 + \epsilon}{\kl{\mu_{\superarmelement}}{\mu_{\superarmbestelement}}} f(\timeidx)  +\frac{\exp\br{-\kl{\phi(\epsilon, \mu_{\superarmbestelement}, \mu_{\superarmelement})}{\mu_{\superarmelement}} K_\timeidx}}{1 - \exp(-\kl{\phi(\epsilon, \mu_{\superarmbestelement}, \mu_{\superarmelement})}{\mu_{\superarmelement}}},
\end{align*}\fi
where $\phi(\epsilon, \mu_{\superarmbestelement}, \mu_{\superarmelement})$ defined on the open interval $]\mu_{\superarmbestelement}, \mu_{\superarmelement}[$ is calculated such that $\kl{\phi(\epsilon, \mu_{\superarmbestelement}, \mu_{\superarmelement})}{\mu_{\superarmbestelement}} = \kl{\mu_{\superarmelement}}{\mu_{\superarmbestelement}}/(1+\epsilon)$.
\begin{proof}[Sketch of proof]
The authors of \cite{Garivier2013} bound the probability of choosing a suboptimal arm based on a \ac{kl}-divergence-based \ac{ucb} for all distributions being part of the exponential family. As we can mirror the probability density function of the exponential distribution at the y-axis and result with a distribution from of the exponential family, we can transfer our \ac{lcb} setting to an equivalent \ac{ucb} setting. Thus, by symmetry, the performance guarantees of the non-combinatorial \ac{kl}-based policy for the exponential family apply.
\end{proof}
\end{lemma}

Starting with a similar approach as in the proof of~\cref{claim:bound_suboptimal_choice} and by help of the relation given in \cref{lemma:kl_suboptimal_total_prob_bound}, we can bound the expectation of the counter $\failcountermax$ as
\ifdouble\begin{align}
    &\E{\failcountermax} \leq
    \sum\limits_{\tmpvar = 1}^{\timeidx} 
    \Pr(\exists\; 1 \leq \element \leq \card{\superarm[\tmpvar]}: \mu_{\superarmelement[\tmpvar]} > \mu_{\superarmbestelement}, \nonumber \\ &\cbshort[\tmpvar]{\superarmelement[\tmpvar]} \leq \cbshort[\tmpvar]{\superarmbestelement}) \! \nonumber \\ %
    &\leq \sum\limits_{\tmpvar = 1}^{\timeidx} 
    \sum\limits_{\element = 1}^{\card{\superarm[\tmpvar]}} \Prob{\cbshort[\tmpvar]{\superarmelement[\tmpvar]} \leq \cbshort[\tmpvar]{\superarmbestelement}, \mu_{\superarmelement[\tmpvar]} > \mu_{\superarmbestelement}} \label{eq_line:bounded_element} \\
    &\leq \sum\limits_{\element = 1}^{\cardmax} \sum\limits_{\tmpvar = 1}^{\timeidx} 
    \Prob{\cbshort[\tmpvar]{\superarmelement[\tmpvar]} \leq \cbshort[\tmpvar]{\superarmbestelement}, \mu_{\superarmelement[\tmpvar]} > \mu_{\superarmbestelement}} \label{eq_line:unbounded_element} \\
    &\leq \sum\limits_{\element = 1}^\cardmax 7 \log(\log(\timeidx)) + \frac{1 + \epsilon}{\klminus{\mu_{\superarmelement}}{\mu_{\superarmbestelement}}} f(\timeidx) \nonumber \\ &+ \frac{\exp\br{-\kl{\phi(\epsilon, \mu_{\superarmbestelement}, \mu_{\tmpset_\element})}{\mu_{\superarmelement}} K_\timeidx}}{1 - \exp(-\kl{\phi(\epsilon, \mu_{\superarmbestelement}, \mu_{\tmpset_\element})}{\mu_{\superarmelement}}} \nonumber \\
    &\leq \cardmax \br{7 \log(\log(\timeidx)) + \frac{1 + \epsilon}{\klmin} f(\timeidx) + Q_\iter(\epsilon)}, \nonumber
\end{align}\else
\begin{align}
    &\E{\failcountermax} \leq
    \sum\limits_{\tmpvar = 1}^{\timeidx} 
    \Pr(\exists\; 1 \leq \element \leq \card{\superarm[\tmpvar]}: \mu_{\superarmelement[\tmpvar]} > \mu_{\superarmbestelement}, \nonumber \cbshort[\tmpvar]{\superarmelement[\tmpvar]} \leq \cbshort[\tmpvar]{\superarmbestelement}) \! \nonumber \\ %
    &\leq \sum\limits_{\tmpvar = 1}^{\timeidx} 
    \sum\limits_{\element = 1}^{\card{\superarm[\tmpvar]}} \Prob{\mu_{\superarmelement[\tmpvar]} > \mu_{\superarmbestelement}, \cbshort[\tmpvar]{\superarmelement[\tmpvar]} \leq \cbshort[\tmpvar]{\superarmbestelement}} \label{eq_line:bounded_element} \\
    &\leq \sum\limits_{\element = 1}^{\cardmax} \sum\limits_{\tmpvar = 1}^{\timeidx} 
    \Prob{\mu_{\superarmelement[\tmpvar]} > \mu_{\superarmbestelement}, \cbshort[\tmpvar]{\superarmelement[\tmpvar]} \leq \cbshort[\tmpvar]{\superarmbestelement}} \label{eq_line:unbounded_element} \\
    &\leq \sum\limits_{\element = 1}^\cardmax 7 \log(\log(\timeidx)) + \frac{1 + \epsilon}{\klminus{\mu_{\superarmelement}}{\mu_{\superarmbestelement}}} f(\timeidx) \nonumber + \frac{\exp\br{-\kl{\phi(\epsilon, \mu_{\superarmbestelement}, \mu_{\tmpset_\element})}{\mu_{\superarmelement}} K_\timeidx}}{1 - \exp(-\kl{\phi(\epsilon, \mu_{\superarmbestelement}, \mu_{\tmpset_\element})}{\mu_{\superarmelement}}} \nonumber \\
    &\leq \cardmax \br{7 \log(\log(\timeidx)) + \frac{1 + \epsilon}{\klmin} f(\timeidx) + Q_\iter(\epsilon)}, \nonumber
\end{align}\fi
where from \eqref{eq_line:bounded_element} to \eqref{eq_line:unbounded_element} we used the convention that $\mathds{1} \{\cbshort[\tmpvar]{\superarmelement[\tmpvar]} \leq \cbshort[\tmpvar]{\superarmbestelement}, \\ \mu_{\superarmelement[\tmpvar]} > \mu_{\superarmbestelement}\} = 0$ if $\element > \card{\superarm[\tmpvar]}$. This concludes the proof.
\end{proof}

\vfill

\end{document}